\documentclass[acmsmall,screen]{acmart}
\usepackage[shortlabels]{enumitem}
\usepackage{fancybox, graphicx}
\usepackage{color}
\usepackage{listings}
\usepackage{tcolorbox}
\usepackage{balance}
\usepackage{tabularx}
\usepackage{multirow}
\usepackage{xspace}
\usepackage{tikz}
\usepackage{textcomp}
\usepackage{booktabs}
\usetikzlibrary{patterns}
\usepackage{url}

\definecolor{darkblue}{rgb}{0.0, 0.0, 0.55}  

\usepackage[normalem]{ulem}
\usepackage[utf8]{inputenc}
\usepackage{amsmath}  
\usepackage{graphicx}
\usepackage{comment}
\usepackage{multirow}
\usepackage{enumitem}
\usepackage{array}
\usepackage{pifont}
\usepackage{xcolor}
\usepackage{balance}
\usepackage{mathtools}
\usepackage{xspace}
\usepackage{url}
\usepackage{eucal}
\usepackage{amsfonts}
\usepackage{color}
\usepackage{booktabs}
\usepackage{bbding}
\usepackage{float}

\usepackage{graphicx}
\usepackage{hyperref}
\usepackage{url}
\usepackage{multirow}
\usepackage{comment}
\usepackage{colortbl}
\usepackage{arydshln}
\usepackage{graphicx}
\usepackage{subcaption}
\usepackage[linewidth=0.1pt]{mdframed}
\usepackage[linesnumbered,ruled,lined]{algorithm2e}

\usepackage{wrapfig,lipsum,booktabs}
\usepackage{subfloat}
\usepackage[export]{adjustbox}%

\usepackage{graphicx}
\usepackage{adjustbox}
\usepackage{forest}
\usepackage{tikz}
\usepackage{pifont}

\usepackage{hyperref}
\usepackage{hyperxmp}

\AtBeginDocument{%
  \providecommand\BibTeX{{%
    \normalfont B\kern-0.5em{\scshape i\kern-0.25em b}\kern-0.8em\TeX}}}

\newcommand{\revised}[1]{\textcolor{black}{#1}}

\pdfpagewidth=\paperwidth
\pdfpageheight=\paperheight

\makeatletter
\newcommand{\showfontsize}{\f@size{} pt}
\newcommand\usemm[1]{%
  \strip@pt\dimexpr0.3514598\dimexpr #1\relax\relax mm%
}
\newcommand\usein[1]{%
  \strip@pt\dimexpr0.013837\dimexpr #1\relax\relax in%
}
\makeatother

\usepackage{graphicx}
\usepackage{hyperref}
\usepackage{url}
\usepackage{multirow}
\usepackage{comment}
\usepackage{colortbl}
\usepackage{graphicx}
\usepackage{tcolorbox}

\usepackage{algpseudocode}  
\usepackage{amsmath}  
\usepackage{caption} 
\usepackage{algpseudocode}
\usepackage{float}
\usepackage{soul}

\usepackage{xcolor}
\usepackage{enumitem}

\usepackage{amsmath}

\definecolor{dark-red}{RGB}{255,0,0}
\definecolor{dark-green}{RGB}{0,200,0}
\definecolor{lightgreen}{rgb}{0.56, 0.93, 0.56} 
\definecolor{lightred}{rgb}{1.0, 0.7, 0.7}

\lstdefinelanguage{java-pretty}
{
  language=java,
  numbers=left,
  frame=shadowbox,
  rulesepcolor= \color{red!20!green!20!blue!20},
  basicstyle=\footnotesize\ttfamily,
  numberstyle=\scriptsize,
  breaklines=true,
  columns=fullflexible,
  xleftmargin=16pt,
  showstringspaces=false,
  keywordstyle=\color{blue}\bfseries,
  stringstyle=\color{javared},
  commentstyle=\color{javagreen},
  morecomment=[s][\color{javadocblue}]{/**}{*/},
}
\colorlet{punct}{red!60!black}
\definecolor{background}{HTML}{EEEEEE}
\definecolor{delim}{RGB}{20,105,176}
\colorlet{numb}{magenta!60!black}
\lstdefinelanguage{json}{
    basicstyle=\normalfont\ttfamily,
    numbers=left,
    numberstyle=\scriptsize,
    stepnumber=1,
    numbersep=8pt,
    showstringspaces=false,
    breaklines=true,
    frame=lines,
    literate=
     *{:}{{{\color{punct}{:}}}}{1}
      {,}{{{\color{punct}{,}}}}{1}
      {\{}{{{\color{delim}{\{}}}}{1}
      {\}}{{{\color{delim}{\}}}}}{1}
      {[}{{{\color{delim}{[}}}}{1}
      {]}{{{\color{delim}{]}}}}{1},
}

\lstdefinelanguage{json-pretty}
{
  language=json,
  numbers=left,
  frame=shadowbox,
  rulesepcolor= \color{red!20!green!20!blue!20},
  basicstyle=\footnotesize\ttfamily,
  numberstyle=\scriptsize,
  breaklines=true,
  columns=fullflexible,
  xleftmargin=16pt,
  showstringspaces=false,
  keywordstyle=\color{blue}\bfseries,
  stringstyle=\color{javared},
  commentstyle=\color{javagreen},
  morecomment=[s][\color{javadocblue}]{/**}{*/},
}

\lstdefinelanguage{diff}{
    basicstyle=\ttfamily\small,
    morecomment=[f][\color{diffstart}]{@@},
    morecomment=[f][\color{javagreen}]{+\ },
    morecomment=[f][\color{javared}]{-\ },
  }

\newcommand{\InputWithSpace}[1]{\bgroup\def\arraystretch{1.15}\input{#1}\egroup}

\setcopyright{acmcopyright}
\copyrightyear{2024}
\acmYear{2024}
\acmDOI{XXXXXXX.XXXXXXX}

\begin{document}

\title{LessLeak-Bench: A First Investigation of Data Leakage in LLMs Across 83 Software Engineering Benchmarks}

\author{Xin Zhou}
\affiliation{%
  \institution{Singapore Management University}
  \country{Singapore}
}
\email{xinzhou.2020@phdcs.smu.edu.sg}

\author{Martin Weyssow}
\affiliation{%
  \institution{Singapore Management University}
  \country{Singapore}
}
\email{mweyssow@smu.edu.sg}

\author{Ratnadira WIDYASARI}
\affiliation{%
  \institution{Singapore Management University}
  \country{Singapore}
}
\email{ratnadiraw.2020@phdcs.smu.edu.sg}

\author{Ting Zhang}
\affiliation{%
  \institution{Singapore Management University}
  \country{Singapore}
}
\email{tingzhang.2019@phdcs.smu.edu.sg}

\author{Junda He}
\affiliation{%
  \institution{Singapore Management University}
  \country{Singapore}
}
\email{jundahe@smu.edu.sg}

\author{Yunbo Lyu}
\affiliation{%
  \institution{Singapore Management University}
  \country{Singapore}
}
\email{yunbolyu@smu.edu.sg}

\author{Jianming Chang}
\email{jianmingchang@seu.edu.cn}
\affiliation{%
  \institution{Southeast University}
  \city{Nanjing}
  \state{JiangSu}
  \country{China}
}

\author{Beiqi Zhang}
\email{zhangbeiqi@whu.edu.cn}
\affiliation{%
  \institution{Wuhan University}
  \city{Wuhan}
  \country{China}
}

\author{Dan Huang}
\affiliation{%
  \institution{Singapore Management University}
  \country{Singapore}
}
\email{dan.huang.2024@phdcs.smu.edu.sg}

\author{David Lo}
\affiliation{%
  \institution{Singapore Management University}
  \country{Singapore}
}
\email{davidlo@smu.edu.sg}

\renewcommand{\shortauthors}{Zhou et al.}

\begin{abstract}
Large Language Models (LLMs) are widely utilized in software engineering (SE) tasks, such as code generation and automated program repair. However, their reliance on extensive and often undisclosed pre-training datasets raises significant concerns about data leakage, where the evaluation benchmark data is unintentionally ``seen'' by LLMs during the model's construction phase.
The data leakage issue could largely undermine the validity of LLM-based research and evaluations. 
Despite the increasing use of LLMs in the SE community, there is no comprehensive study that assesses the extent of data leakage in SE benchmarks for LLMs yet.
To address this gap, this paper presents the first large-scale analysis of data leakage in 83 SE benchmarks concerning LLMs. 
We systematically investigated whether, and to what extent, popular SE benchmark datasets were included in a LLM's pre-training data. 
Our approach involved using an efficient near-duplicate data detection algorithm, MinHash+LSH, to identify potential duplicate pairs between the SE benchmarks and LLM's pre-training dataset. Subsequently, we conducted extensive manual labeling on these potential duplicates to identify true duplicates.
Those true duplicates reveal and confirm the data leakage of SE benchmarks. 
Our results show that in general, data leakage in SE benchmarks is minimal, with average leakage ratios of only 4.8\%, 2.8\%, and 0.7\% for Python, Java, and C/C++ benchmarks, respectively. However, some benchmarks exhibit relatively higher leakage ratios, which raises concerns about their bias in evaluation. For instance, QuixBugs and BigCloneBench have leakage ratios of 100.0\% and 55.7\%, respectively.
Furthermore, we observe that data leakage has a substantial impact on LLM evaluation. On the APPS benchmark, StarCoder-7B achieves a Pass@1 score that is 4.9 times higher on leaked samples than on non-leaked samples, highlighting how leaked benchmark data can lead to inflated metrics.
We also identify key causes of high data leakage, such as the direct inclusion of benchmark data in pre-training datasets and the use of coding platforms like LeetCode for benchmark construction.
To address the data leakage, we introduce \textbf{LessLeak-Bench}, a new benchmark that removes leaked samples from the 83 SE benchmarks, enabling more reliable LLM evaluations in future research.
Our study enhances the understanding of data leakage in SE benchmarks and provides valuable insights for future research involving LLMs in SE.

\end{abstract}

\maketitle

\newcommand{\XSpace}[1]{}

\makeatletter
\newenvironment{btHighlight}[1][]
{\begingroup\tikzset{bt@Highlight@par/.style={#1}}\begin{lrbox}{\@tempboxa}}
{\end{lrbox}\bt@HL@box[bt@Highlight@par]{\@tempboxa}\endgroup}

\newcommand\btHL[1][]{%
  \begin{btHighlight}[#1]\bgroup\aftergroup\bt@HL@endenv%
}
\def\bt@HL@endenv{%
  \end{btHighlight}%
  \egroup
}
\newcommand{\bt@HL@box}[2][]{%
  \tikz[#1]{%
    \pgfpathrectangle{\pgfpoint{1pt}{0pt}}{\pgfpoint{\wd #2}{\ht #2}}%
    \pgfusepath{use as bounding box}%
    \node[anchor=base west, fill=yellow!30,outer sep=0pt,inner xsep=1pt, inner ysep=0pt, rounded corners=0pt, minimum height=\ht\strutbox+1pt,#1]{\raisebox{1pt}{\strut}\strut\usebox{#2}};
  }%
}
\makeatother
\lstdefinestyle{Highlight}{
    moredelim=**[is][\btHL]{`}{`},
    moredelim=**[is][{\btHL[fill=orange!50]}]{´}{´},
    moredelim=**[is][{\btHL[fill=red!50]}]{@}{@},
}

\newcommand{\wholedata}{{\textit{complete data}}}
\newcommand{\singledata}{{\textit{single-line data}}}

\section{Introduction}
Recently, Large Language Models (LLMs) have been extensively applied to software engineering (SE) tasks~\cite{10.1145/3695988}, such as code generation~\cite{DBLP:conf/icse/YuSRZZMLLWX24} and automated program repair~\cite{DBLP:conf/icse/XiaWZ23}, leading to notable advancements in the SE domain. The remarkable capabilities of LLMs arise from the extensive knowledge within their large pre-training data, which encompasses various data types, including code, texts, or other modalities.

Despite their effectiveness, many advanced LLM providers—both open-source and commercial—often do not disclose their pre-training datasets~\cite{llama3, gpt-3.5-turbo}. This lack of transparency raises a critical concern in evaluating LLM-related approaches: the risk of data leakage. Data leakage occurs when LLMs are exposed to SE benchmark datasets during their pre-training phase, compromising the evaluation's validity.
The impact of data leakage is twofold. First, it complicates the assessment of whether the notable performance of LLM-based approaches arises from true innovation or inflated effectiveness metrics due to prior exposure to SE benchmark data. Second, it leads to an unfair comparison between LLM-based methods and non-learning-based techniques, such as traditional program analysis approaches, which do not rely on training data and have no opportunity to learn from leaked data.
The data leakage issue in LLMs is becoming increasingly recognized and significant due to LLMs' growing usage~\cite{lópez2024interdatasetcodeduplicationdata}. However, there is a lack of comprehensive studies investigating whether, and to what extent, SE benchmarks have been leaked into the pre-training data of LLMs.

To address this gap, we conduct the first large-scale investigation into data leakage of SE benchmarks. Our study investigates \textbf{83} diverse SE benchmarks\footnote{If a benchmark has multiple variants, we treat each variant as a separate dataset or benchmark.} across three widely used programming languages: Java, C/C++, and Python. Our study covers various SE tasks, such as code generation, code editing, program repair, fault localization, API recommendation, code translation, clone detection, test generation, vulnerability detection, vulnerability repair, debugging, log statement generation, and automatic code review. 
By conducting an extensive analysis of these benchmarks, we aim to understand the comprehensive landscape of data leakage issues within the SE domain. 
Regarding the studied LLMs, we focus on data leakage within the \textbf{StarCoder}~\cite{starcoder_one} family. We choose StarCoder for three key reasons: (1) we require fully open-source LLMs—with publicly available pre-training data—to identify concrete evidence of data leakage; (2) StarCoder demonstrates competitive performance among fully open-source models; and (3) it serves as the foundation for derivative LLMs such as WizardCoder~\cite{luo2023wizardcoder}, OctoPack~\cite{muennighoff2023octopack}, CodeShell~\cite{xie2024codeshell}, and DeepSeek-Coder~\cite{deepseekcoder}.

\begin{figure*}[t] 
    \centering
    \includegraphics[width=14cm]{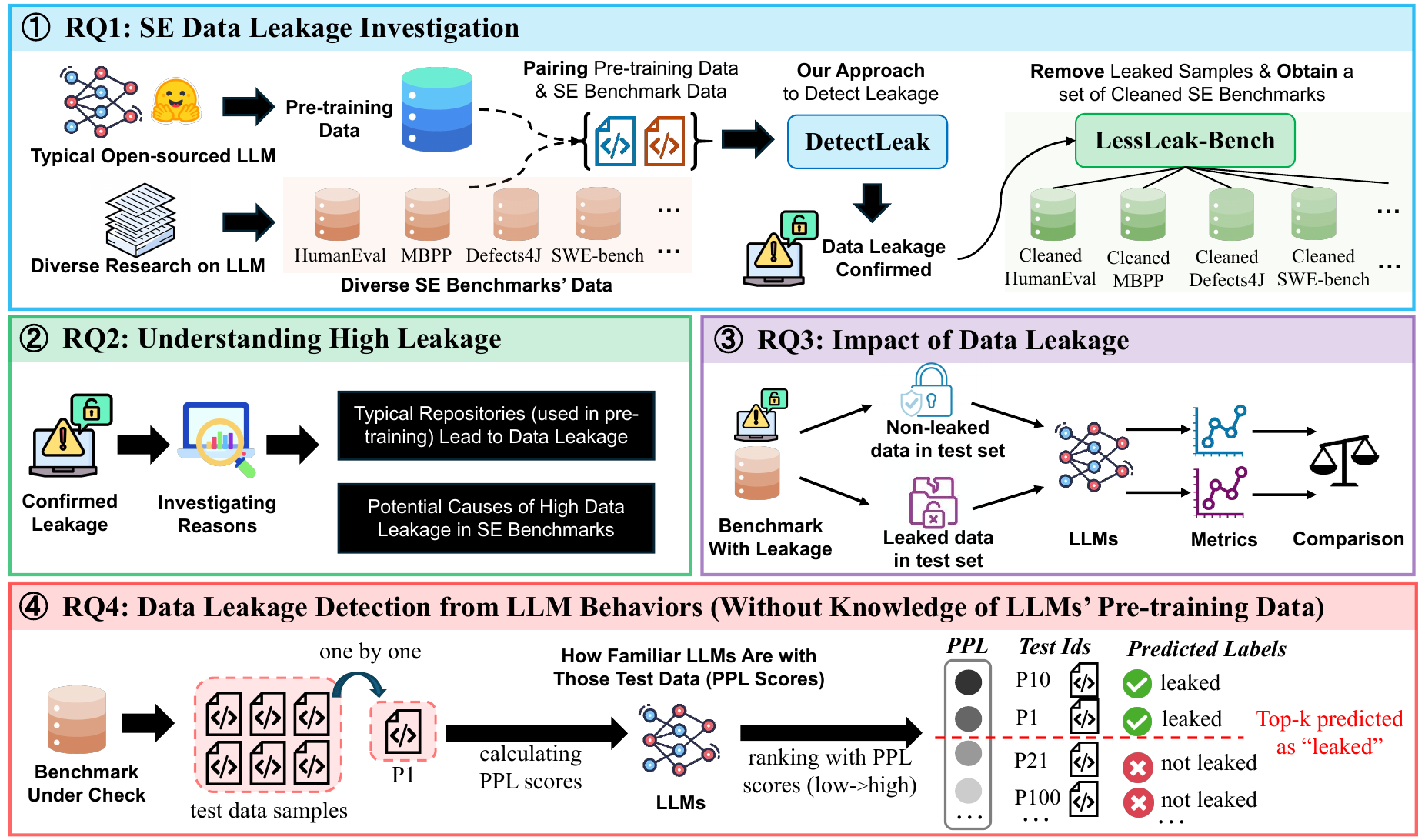} 
    \vspace{-0.5cm}
    \caption{Overview of Our Study} 
    \vspace{-0.3cm}
    \label{fig:framework}
\end{figure*}

Studying data leakage in SE benchmarks requires detecting duplicate data between the SE benchmark datasets and the pre-training data of LLMs. To rigorously examine data leakage in SE benchmarks, we proposed a multi-phase approach named \textbf{DetectLeak}, which combines automated techniques with manual labeling to identify overlaps between LLM pre-training data and SE benchmark datasets.
Initially, DetectLeak employs an automated near-duplication detection tool namely MinHash+LSH, to identify potential duplicate pairs. This involves comparing approximately 1.7 trillion pairs of LLM pre-training and SE benchmark data.
Next, several experienced developers collaboratively label the potential duplicate pairs flagged by the automated tool, to accurately identify true duplicates and thus confirm the existence of data leakage of SE benchmark data.

By applying DetectLeak, we identified and confirmed 606, 816, and 108 leaked data samples in Python, Java, and C/C++ SE benchmarks, respectively.
In general, data leakage in SE benchmarks is minimal, with average leakage ratios of only 4.8\%, 2.8\%, and 0.7\% for Python, Java, and C/C++ benchmarks, respectively. However, some benchmarks exhibit relatively high leakage ratios, which raises concerns about evaluation bias. For instance, QuixBugs~\cite{QuixBugs} has the highest leakage rate at 100.0\%, followed by BigCloneBench~\cite{BigCloneBench} with a significant leakage rate of 55.7\%. Other notable benchmarks, such as APPS~\cite{hendrycks2021measuring}, BugsInPy~\cite{BugsInPy}, SWE-Bench-verified~\cite{SWE-bench-verified}, and CodeEditorBench~\cite{CodeEditorBench}, also show leakage rates ranging from 8–10\%.

To address the identified data leakage, we introduce \textbf{LessLeak-Bench}, a new benchmark that provides cleaned versions of 83 studied SE benchmarks with all identified leaked samples removed. We believe LessLeak-Bench will enable more reliable evaluations of LLMs in future research.

In this work, we structure our study by answering the following research
questions, with an overview illustrated in Figure~\ref{fig:framework}:
\begin{itemize}[leftmargin=*]
\item[-]  \textbf{RQ1. To what extent does data leakage exist in the studied SE benchmarks?}
\item[-]  \textbf{RQ2. What factors contribute to high leakage rates in SE benchmarks?} 
\item[-]  \textbf{RQ3. How does the leakage of benchmark data affect the effectiveness of LLMs?}
\item[-]  \textbf{RQ4. How effective is the automated metric in detecting data leakage when lacking access to LLM pre-training data?}

\end{itemize}

Based on our results, we present the following insightful findings:
1) Overall, data leakage in SE benchmarks is minimal; however, some benchmarks show relatively higher leakage ratios, such as QuixBugs (100\%), BigCloneBench(55.7\%), APPS (10.8\%), and SWE-Bench-verified (10.6\%). Researchers should be cautious of potential data leakage in the benchmarks they use.
2) We identified several potential causes of high data leakage, such as the direct inclusion of benchmark data in the pre-training dataset or overlap between the repositories used to create the benchmarks and those included in the pre-training dataset.
3) Data leakage has a significant impact on the evaluation of LLMs. For example, StarCoder-7b achieves a Pass@1 score that is 4.9 times higher on the leaked samples than on the non-leaked samples on the APPS benchmark. This underscores how the presence of leaked benchmark samples can lead to significantly inflated metrics, emphasizing the critical need to address the data leakage issue in SE benchmarks to ensure fair evaluation.
4) Developing more advanced methods to effectively detect leaked benchmark samples, especially without access to LLM pre-training data, remains a significant challenge and a promising direction for future research.

In summary, our contributions are as follows:
\begin{itemize}[leftmargin=*]
\item To the best of our knowledge, this is the first extensive investigation into data leakage across a comprehensive set of 83 SE benchmarks. Our study spans three widely used programming languages—Java, C, and Python—and covers diverse popular tasks in software engineering.
\item We reveal and confirm that data leakage is notably prevalent in certain SE tasks/benchmarks while less significant in others. Additionally, we analyze the potential reasons that contribute to the relatively high leakage and we offer actionable recommendations based on our findings.
\item We developed a framework namely DetectLeak, which can identify and quantify the extent of data leakage. In addition, we offer a cleaned set of SE benchmarks, LessLeak-bench, which removes identified data leaks across diverse Java, C/C++, and Python SE benchmarks. 
\end{itemize}

The remainder of this paper is organized as follows. Section \ref{section2} formalizes the data leakage problem. In Section \ref{section3}, we describe the details of our proposed approach to detect data leakage.
Section \ref{section4} explains our experimental settings. 
Section \ref{section5} presents experimental results and the answers to research questions. 
We present the discussion and related work in Section \ref{section6} and Section \ref{section7}, respectively. Finally, Section \ref{section8} concludes the paper and highlights directions for future research.

\section{Background}\label{section2}

\subsection{Data Leakage Definition}

Data leakage (also called Data Contamination) in LLMs refers to the unintentional inclusion of evaluation data during the model's construction phase (e.g., pre-training or training)~\cite{balloccu2024leak}. This issue can lead to an inaccurate assessment of a model's true capabilities. To formally define data leakage, we denote \( D_{construct} \) as the pre-training/training dataset used for model construction and \( D_{eval} \) as the evaluation dataset. Data leakage occurs if either of the following two conditions is met:

\vspace{0.1cm}
\noindent
1. \textbf{Exact Leakage.} At least one evaluation sample exactly matches a data sample in the pre-training or training set: \[D_{eval} \cap D_{construct} \neq \emptyset \]

\vspace{0.1cm}
\noindent
2. \textbf{Semantic Leakage.} At least one evaluation sample is semantically equivalent to a data sample in the pre-training or training set:
\[
   \exists x_{eval} \in D_{eval}, \exists x_{construct} \in D_{construct} \colon
   f(x_{eval}, x_{construct}) = 1
\]
Here, \( f(x_{\text{eval}}, x_{\text{construct}}) \) is a function that measures the semantic equivalence between two data samples. If the samples are semantically equivalent, the function returns 1; otherwise, it returns 0. This condition indicates that at least one evaluation sample is semantically equivalent to a pre-training/training sample.

\vspace{0.2cm}
\noindent\textbf{Data Leakage Regarding LLMs.}
For approaches based on LLMs, leakage can occur in two distinct phases: (1) during pre-training, where evaluation data is inadvertently included in the pre-training dataset, and (2) during fine-tuning or prompting, where evaluation data is accidentally incorporated into the training set.
Detecting the second type of data leakage, which occurs between training and evaluation data, is straightforward. Since the training data is usually available, researchers can easily check for any overlap with the evaluation data (or test set). In contrast, identifying the first type of data leakage, which involves the relationship between pre-training and evaluation data, presents significant challenges. This difficulty arises from two main factors: 1) pre-training datasets for LLMs are often undisclosed, making it difficult to ascertain their contents, and 2) the sheer size of pre-training datasets renders it computationally expensive to examine all data for potential overlaps.
In this study, we focus on investigating data leakage between the pre-training dataset of LLMs and diverse SE benchmarks that serve as evaluation data for various SE tasks.

\subsection{The Relationship Between Data Leakage Detection and Code Clone Detection}
\revised{
Data leakage detection and code clone detection both seek to identify duplicate data pairs but serve distinct purposes. Clone detection focuses on identifying semantically equivalent or identical code snippets, aiming to reduce redundancy and improve maintainability. In contrast, data leakage detection examines the overlap between benchmark data and an LLM's pre-training corpus to determine whether the model has encountered test data during pre-training or training. Furthermore, while code clone detection is usually limited to duplicate source code, data leakage detection applies to a wider array of data formats, including code, natural language texts, and structured data. 
Finally, the actions taken after detection also differ. Code clone detection typically leads to the removal or merging of duplicates to optimize the codebase, while data leakage detection focuses on eliminating overlaps between the benchmark dataset and pre-training/training data to preserve the validity of LLM evaluations.
}

\section{Methodology}\label{section3}

\begin{figure}[t] 
    \centering
    \includegraphics[width=14cm]{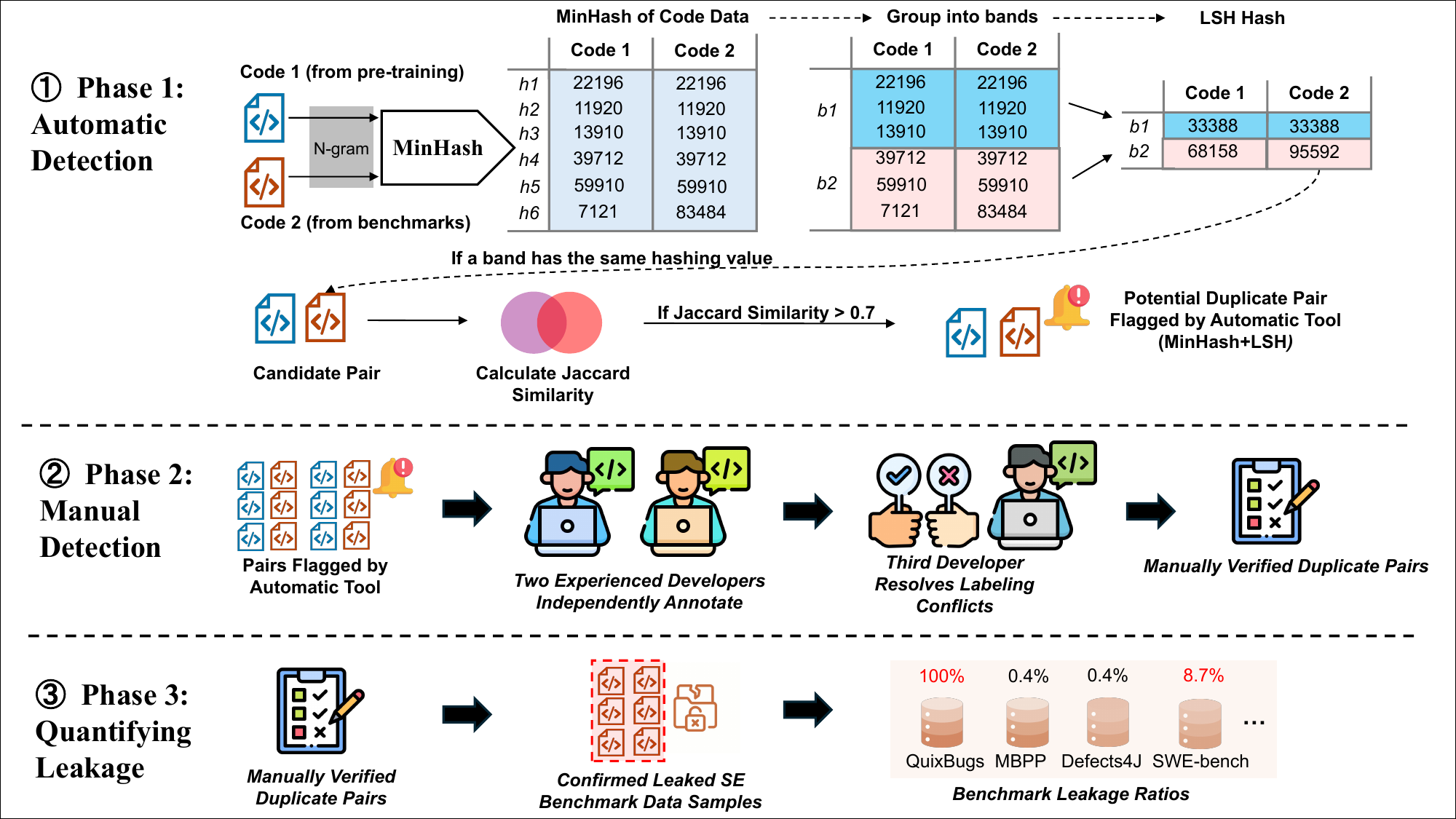} 
     \vspace{-0.3cm}
    \caption{Our Data Leakage Detection Pipeline: DetectLeak} 
     \vspace{-0.6cm}
    \label{fig:method_minhash}
\end{figure}

This section describes our framework, \textbf{DetectLeak}, for detecting duplicate data between SE benchmarks and the LLM’s pre-training corpus.
Figure~\ref{fig:method_minhash} presents our DetectLeak framework, which consists of three main steps:

\vspace{0.1cm}
\noindent
\textbf{Step 1: Automatic Detection of Potential Duplicates (\textcircled{1} of Figure~\ref{fig:method_minhash}):} Given an LLM pre-training corpus and an SE benchmark, we use an automated near-duplicate detection tool to identify potential duplicate pairs between the SE benchmark data and the LLM's training data.

\vspace{0.1cm}
\noindent
\textbf{Step 2: Manual Verification on Potential Duplicates (\textcircled{2} of Figure~\ref{fig:method_minhash}:)} Experienced developers collaboratively review and annotate the potential duplicates identified in Step 1. This manual validation ensures the accurate identification of true duplicates.

\vspace{0.1cm}
\noindent
\textbf{Step 3: Quantifying the Extent of Data Leakage (\textcircled{3} of Figure~\ref{fig:method_minhash}):}  In this step, we first identify the leaked benchmark data samples from the manually verified duplicate pairs between the SE benchmark data and the LLM's training data. Then, we quantify the extent of data leakage for each studied SE benchmark and compute their corresponding leakage ratios.

\subsection{Automated Data Leakage Detection}

While automated methods may not achieve the precision of manual labeling, manually reviewing extensive pre-training datasets containing millions of samples is both impractical and cost-prohibitive. Therefore, in the first phase of DetectLeak, we adopt a cost-effective and efficient automated approach to detect potential data leakage. For instance, in this study, we conduct the comparison of 1.7 trillion pairs of LLM pre-training and SE benchmark data,  based on our studied LLM and SE benchmarks, which will be introduced in   Sections~\ref{llm_selection} and~\ref{benchmark_selection}. The sheer scale of this computational effort highlights the challenge posed by the extensive size of LLM pre-training datasets, emphasizing the complexity and resource-intensive nature of studying data leakage regarding LLMs.
This also supports our decision to use the efficient automated tool as the initial step before manual labeling.

\vspace{0.2cm}
\noindent\textbf{Automated Duplicate Detection Tool Selection.}
We specifically choose \textbf{MinHash+LSH}~\cite{SantaCoder, CodeParrot}, a scalable technique for identifying potential duplicates in large datasets. This tool has been widely adopted by many LLMs such as Qwen~\cite{bai2023qwen}, CodeParrot~\cite{CodeParrot}, SantaCoder~\cite{SantaCoder}, StarCoder~\cite{starcoder_one}, and StarCoder v2~\cite{starcoder2} to detect and filter potential duplicate data within their pre-training corpora. 
While the authors of these LLMs use MinHash+LSH for filtering duplicates in their pre-training datasets, we adopt it for a different purpose: detecting potential duplicates between LLM pre-training data and SE benchmark datasets.
We also considered other automated tools for duplicate data detection except for MinHash+LSH but decided against using them for several reasons. 

First, Exact Match is effective for detecting exact duplicates but fails to identify semantic duplicates (i.e., those that are semantically equivalent but not identical), which are crucial for our analysis. 
Second, while many code clone detection approaches are available in the literature~\cite{DBLP:journals/jss/NasrabadiPRRE23}, many effective tools are learning-based and rely on training with data that closely matches the distribution of the test set. These tools are also typically evaluated on Java-based benchmarks, (e.g., BigCloneBench~\cite{BigCloneBench}), and focus primarily on code snippets. 
In contrast, our study spans three different programming languages, and we lack training data for duplicates between SE benchmarks and LLM pre-training data. 
Additionally, benchmarks such as SWE-Bench~\cite{SWE-bench} contain not only code but also other types of content, such as GitHub issue reports (similar to bug reports), which require a tool that can handle both code and texts.
MinHash+LSH, on the other hand, does not require training data, supports multiple programming languages, and is capable of handling both code and text. These features make it a more suitable choice for our study.
Furthermore, we have observed the widespread use of MinHash+LSH in deduplicating LLM pre-training data for models such as Qwen, CodeParrot, SantaCoder, StarCoder, and StarCoder v2. However, we have not noticed code clone detection tools applied to LLM' huge pre-training data to detect duplicates yet. For these reasons, we chose to follow the established practice of many LLM developers to select MinHash+LSH for our analysis.

\vspace{0.2cm}
\noindent\textbf{MinHash+LSH Method Details.}
As shown in \textcircled{1} of Figure~\ref{fig:method_minhash}, the method consists of several steps to detect potential duplicate pairs. For a given pair of data samples, the method first decomposes each sample into n-gram tokens. Then, the MinHash~\cite{DBLP:conf/cpm/Broder00} module is applied to generate hashes for each sample. Next, Locality-Sensitive Hashing (LSH)~\cite{LSH} is used to group these hashes into bands and hash the bands. By comparing the hashes of bands, the method identifies similar data samples: if a pair of data samples share the same band hash, they are considered candidate duplicate pairs.
For each candidate pair, the method calculates the Jaccard similarity coefficient~\cite{Jaccard_index} between the two data samples in this pair. If the coefficient exceeds the threshold, the pair is flagged as a potential duplicate by the automated tool. In this study, we set the n-gram size to 2 and the Jaccard similarity coefficient threshold to 0.7, based on initial small-scale trials.

\subsection{Manual Labeling of Potential Duplicates}

To improve detection accuracy, our approach includes a manual verification step of potential duplicate pairs flagged by the automated tool. 
This large-scale manual labeling process ensures accurate classification of duplicate pairs, confirming the presence of leaked SE benchmark data samples and revealing the extent of data leakage in software engineering benchmarks.
This annotation process is performed by a team of eight experienced data annotators, including three post-doctoral researchers, four PhD students, and one Master's student, each with a minimum of four years of experience in programming. Figure~\ref{fig:method_minhash} (\textcircled{2}) illustrates the annotation process.

\vspace{0.1cm}
\noindent \textbf{Annotation Scheme.} 
Annotators are tasked with categorizing potential duplicate pairs into one of the following four classifications:
\noindent
\begin{itemize} [left=0pt]
\setlength{\itemsep}{0pt} 
\setlength{\parskip}{0pt} 
\setlength{\parsep}{0pt}  
\item [-] \textit{Not Related}: The two data samples in the pair are dissimilar. 
\item [-] \textit{Related but Not Duplicates}:  
The two data samples address similar tasks but are designed for distinct purposes.
\item [-] \textit{Semantically Equivalent}: The code may feature renamed identifiers (e.g., variables, constants) while preserving the same logic. The structure may remain largely unchanged with minor additions or deletions. Essentially, the code achieves the same functionality but uses different syntax or implementation. 
\item [-] \textit{Exact Copies}: The pairs are identical except for variations in formatting or comments. 
\end{itemize}
\noindent
In this categorization scheme, the first two categories represent the ``\textit{Non-duplicate}'' class, while the latter two correspond to the ``\textit{Duplicate}'' class. We adopt this four-class categorization to gain more detailed insights from annotators, compared to a simple binary classification (non-duplicate or duplicate).

\vspace{0.1cm}
\noindent \textbf{Annotation Process.} 
In our study, the automated tool MinHash+LSH identified 6,643 potential duplicate pairs, based on the selected LLM and SE benchmarks, which will be discussed in Sections~\ref{llm_selection} and~\ref{benchmark_selection}. Given the large volume of data to label, it is impractical for all annotators to review every pair.
To address this, each potential duplicate pair is independently reviewed by two annotators, who evaluate the similarity between the data samples and assign one of the four annotation categories (ranging from ``\textit{Not Related}'' to ``\textit{Exact Copies}'').
After completing the labeling, we calculate the Cohen's Kappa Score~\cite{Cohens_kappa} between the two annotators based on their labels. The average Cohen's Kappa Score is 0.9424, indicating almost perfect agreement~\cite{landis1977measurement}. Notably, when we simplify the four-class categorization (ranging from ``\textit{Not Related}'' to ``\textit{Exact Copies}'') into a binary classification (``\textit{Duplicate}'' or ``\textit{Non-duplicate}''), the Cohen's Kappa Score increases to 0.9591, reflecting even higher agreement on determining whether pairs are duplicates or not.
Although the agreement between the two annotators is high, there are still some pairs where they disagree on the label. In such cases, a third annotator will resolve the conflict by reviewing the labels assigned by the previous two annotators and re-assigning the most appropriate label for the pair.
Finally, after applying our manual labeling method, we identified 1,950 real duplicate pairs and 4,741 non-duplicate pairs out of the 6,691 pairs flagged by the automated tool. These statistics are based on the selected LLM and SE benchmarks, which will be later discussed in Sections~\ref{llm_selection} and~\ref{benchmark_selection}.

\subsection{Quantifying the Extent of Data Leakage}
In the previous steps, we identified the actual duplicate pairs between the SE benchmark data and the LLM’s training data. Using these duplicate pairs, we can easily identify the leaked benchmark data samples. If a pair is marked as a duplicate, the SE benchmark data sample in that pair is considered leaked, as it has been found to have a semantically equivalent or identical counterpart in the LLM’s pre-training data.

Using the confirmed leaked samples, we quantify the extent of data leakage for each studied SE benchmark. To achieve this, we employ two simple and clear metrics: leakage count and leakage ratio. These metrics are defined as follows:
(1) Leakage Count: the total number of leaked benchmark data samples, i.e., $N_{\text{\textit{leak}}}$ ; (2) Leakage Ratio: the proportion of leaked data samples within the benchmark, calculated as: $N_{\text{\textit{leak}}}/N_{\text{\textit{total}}}$.
Here, \(N_{\text{\textit{leak}}}\) denotes the number of identified leaked benchmark samples, and \(N_{\text{\textit{total}}}\) represents the total number of data samples in the benchmark.
A higher leakage count or ratio indicates a more significant degree of data leakage within the benchmark, raising concerns about its reliability for evaluating LLMs.

\subsection{LessLeak-Bench}
After identifying leaked samples in SE benchmarks using our DetectLeak approach, we introduce \textbf{LessLeak-Bench} to mitigate data leakage risks. Specifically, we remove all identified leaked samples from each studied SE benchmark, producing cleaned versions that are free from known data leakage. This curated collection, LessLeak-Bench, serves as a more reliable and comprehensive benchmark for evaluating LLMs across diverse SE tasks.

\section{Experimental Setup}\label{section4}
In this section, we discuss the selection process for the studied LLM and SE benchmarks. Additionally, we outline the implementation details and define the research questions in this study.

\vspace{-0.2cm}
\subsection{LLM Selection} 
\label{llm_selection}

 \textbf{Selection Criteria.} Selecting an appropriate target LLM was a crucial step in this study due to the extensive manual annotation involved, which makes repeated experiments across multiple models infeasible. To guide our selection, we established the following selection criteria:

\noindent
\begin{enumerate}[left=0pt]
    \item \textit{Full Open-Source Availability.} The selected LLM must be fully open-source, encompassing both model parameters and pre-training data. However, since re-training the model is outside the scope of our study, open-source pre-training algorithms and scripts are not required.
    \item \textit{High Effectiveness.} The model should demonstrate strong effectiveness on widely used SE benchmarks, such as HumanEval~\cite{chen2021evaluating} and MBPP~\cite{MBPP_1}.
    \item \textit{Influence and Adoption.} 
We prioritized models with significant influence, particularly those that have inspired or laid the groundwork for the development of newer/better LLMs.
\end{enumerate}

\begin{table*}[t]
\caption{The Overview of Large Language Models (LLMs) for Code.}
 \vspace{-0.2cm}
\label{tab:model_selection}
\centering 
\scalebox{0.75}{ 
\rotatebox{0}{
\begin{tabular}{lllcccrr}
\hline
\textbf{Model}                                             & \textbf{Institution} & \textbf{Size}                                                                                                                              & \textbf{Date} & \textbf{\begin{tabular}[c]{@{}c@{}}Open Source\\ (Model)\end{tabular}} & \textbf{\begin{tabular}[c]{@{}c@{}}Open Source\\ (Data)\end{tabular}} & \multicolumn{1}{c}{\textbf{\begin{tabular}[c]{@{}c@{}}HumanEval \\ (pass@1)\end{tabular}}} & \multicolumn{1}{c}{\textbf{\begin{tabular}[c]{@{}c@{}}MBPP\\ (pass@1)\end{tabular}}} \\ \hline

GPT-C~\cite{svyatkovskiy2020intellicode}  & Microsoft            & 366M                                                                                                                                       & 2020-05       &                                                                        & \multicolumn{1}{l}{}                                                  & -                                                                                          & -                                                                                    \\ 
\rowcolor{lightgreen!50}  CodeBERT~\cite{feng2020codebert}            & Microsoft            & 124M                                                                                                                                       & 2020-09       & \CheckmarkBold                                          & \CheckmarkBold                                         & -                                                                                          & -                                                                                    \\
\rowcolor{lightgreen!50}  GraphCodeBERT~\cite{guo2020graphcodebert}            & Microsoft            & 124M                                                                                                                                       & 2021-02       & \CheckmarkBold                                          & \CheckmarkBold                                         & -                                                                                          & -                                                                                    \\
\rowcolor{lightgreen!50} 
CodeGPT~\cite{lu2021codexglue}            & Microsoft            & 124M                                                                                                                                       & 2021-02       & \CheckmarkBold                                          & \CheckmarkBold                                         & -                                                                                          & -                                                                                    \\ \rowcolor{lightgreen!50} 
GPT-Neo~\cite{gpt-neo}                     & EleutherAI           & 125M, 1.3B, 2.7B
                                                                                       & 2021-03       & \CheckmarkBold                                          & \CheckmarkBold                                         & 6.41                                                                                       & 5.89   \\ 
\rowcolor{lightgreen!50}  PLBART\cite{ahmad2021unified} & UCLA & 140M &	2021-03 & \CheckmarkBold   & \CheckmarkBold & - & -\\\rowcolor{lightgreen!50} 
GPT-J~\cite{gpt-j}                        & EleutherAI           & 6B                                                                                                                                         & 2021-05       & \CheckmarkBold                                          & \CheckmarkBold                                         & 11.62                                                                                      & 11.30                                                                                \\ 
Codex~\cite{chen2021evaluating}           & OpenAI               & 12M-12B
 & 2021-07       &                                                                        & \multicolumn{1}{l}{}                                                  & 28.81                                                                                      & -                                                                                    \\
\rowcolor{lightgreen!50}  CodeT5 \cite{wang2021codet5} & Salesforce & 60M, 220M, 770M & 2021-09 & \CheckmarkBold & \CheckmarkBold & - & - \\
\rowcolor{lightgreen!50}  CodeParrot~\cite{tunstall2022natural}     & Hugging Face         & 110M, 1.5B                                                                                                                                 & 2021-11       & \CheckmarkBold                                          & \CheckmarkBold                                         & 3.99                                                                                       & 1.29                                                                                 \\
\rowcolor{lightgreen!50}  PolyCoder~\cite{xu2022systematic}         & CMU                  & 160M, 400M, 2.7B                                                                                                                           & 2022-02       & \CheckmarkBold                                          & \CheckmarkBold                                         & 5.59                                                                                       & 4.39                                                                                 \\
\rowcolor{lightgreen!50}  CodeGen~\cite{nijkamp2022codegen}         & Salesforce           & 350M-16.1B                                                 & 2022-03       & \CheckmarkBold                                          & \CheckmarkBold                                         & 29.28                                                                                      & 35.28                                                                                \\
\rowcolor{lightgreen!50}  GPT-NeoX~\cite{black2022gpt}              & EleutherAI           & 20B                                                                                                                                        & 2022-04       & \CheckmarkBold                                          & \CheckmarkBold                                         & 15.4                                                                                       & -                                                                                    \\
PaLM-Coder~\cite{chowdhery2023palm}       & Google               & 8B, 62B, 540B                                                                                                                              & 2022-04       &                                                                        & \multicolumn{1}{l}{}                                                  & 36                                                                                         & 47                                                                                   \\
\rowcolor{lightgreen!50}  InCoder~\cite{fried2022incoder}           & Meta                 & 1.3B, 6.7B                                                                                                                                 & 2022-04       & \CheckmarkBold                                          & \CheckmarkBold                                         & 15.2                                                                                       & 21.3                                                                                 \\
PanGu-Coder~\cite{christopoulou2022pangu} & Huawei               & 317M, 2.6B                                                                                                                                 & 2022-07       &                                                                        & \multicolumn{1}{l}{}                                                  & 23.78                                                                                      & 23.0                                                                                 \\
PyCodeGPT~\cite{zan2022cert}              & Microsoft            & 110M                                                                                                                                       & 2022-06       & \CheckmarkBold                                          & \multicolumn{1}{l}{}                                                  & 8.33                                                                                       & 9.39                                                                                 \\
\rowcolor{lightgreen!50}  CodeGeeX~\cite{zheng2023codegeex}         & Tsinghua             & 13B                                                                                                                                        & 2022-09       & \CheckmarkBold                                          & \CheckmarkBold                                         & 22.89                                                                                      & 24.4                                                                                 \\
\rowcolor{lightgreen!50}  BLOOM~\cite{le2023bloom}                  & BigScience           & 176B                                                                                                                                       & 2022-11       & \CheckmarkBold                                          & \CheckmarkBold                                         & 15.52                                                                                      & -                                                                                    \\
GPT-3.5-Turbo~\cite{gpt-3.5-turbo}        & OpenAI               & -                                                                                                                                          & 2022-11       &                                                                        & \multicolumn{1}{l}{}                                                  & 76.2                                                                                       & 52.2                                                                                 \\
\rowcolor{lightgreen!50}  SantaCoder~\cite{allal2023santacoder}     & Hugging Face         & 1.1B                                                                                                                                       & 2022-12       & \CheckmarkBold                                          & \CheckmarkBold                                         & 18                                                                                         & 3.65                                                                                 \\
GPT-4~\cite{gpt4}                & OpenAI               & -                                                                                                                                          & 2023-03       &                                                                        & \multicolumn{1}{l}{}                                                  & 84.1                                                                                       & -                                                                                    \\
\rowcolor{lightgreen!50}  CodeGen2~\cite{nijkamp2023codegen2}       & Salesforce           & 1B-16B                                                                                                                          & 2023-05       & \CheckmarkBold                                          & \CheckmarkBold                                         & 20.46                                                                                      & -                                                                                    \\
\rowcolor{lightred!60}  StarCoder~\cite{starcoder_one}          & Hugging Face         & 15.5B                                                                                                                                      & 2023-05       & \CheckmarkBold                                          & \CheckmarkBold                                         & 33.60                                                                                      & \textcolor{black}{\textbf{52.7}}                                                                                 \\
PanGu-Coder2~\cite{shen2023pangu}         & Huawei               & 15B                                                                                                                                        & 2023-07       &                                                                        & \multicolumn{1}{l}{}                                                  & 61.64                                                                                      & -                                                                                    \\
Llama 2~\cite{llama2}          & Meta                 & 7B, 13B, 70B                                                                                                                               & 2023-07       & \CheckmarkBold                                          & \multicolumn{1}{l}{}                                                  & -                                                                                          & 45.4                                                                                 \\
Code Llama~\cite{roziere2023code}         & Meta                 & 7B, 13B, 34B                                                                                                                               & 2023-08       & \CheckmarkBold                                          & \multicolumn{1}{l}{}                                                  & 48.8                                                                                       & 55                                                                                   \\
phi-1.5~\cite{li2023textbooks}            & Microsoft            & 1.3B                                                                                                                                       & 2023-09       & \CheckmarkBold                                          & \multicolumn{1}{l}{}                                                  & 41.4                                                                                       & 43.5                                                                                 \\
phi-2~\cite{phi-2}                        & Microsoft            & 2.7B                                                                                                                                       & 2023-12       & \CheckmarkBold                                          & \multicolumn{1}{l}{}                                                  & 49.4                                                                                       & 64                                                                                   \\
DeepSeek-Coder~\cite{deepseekcoder}     & DeepSeek             & 1.3B, 6.7B, 33B                                                                                                                            & 2023-11       & \CheckmarkBold                                          & \multicolumn{1}{l}{}                                                  & 56.1                                                                                       & 66                                                                                   \\
\rowcolor{lightgreen!60}  StarCoder2~\cite{starcoder2}   & Hugging Face         & 15B                                                                                                                                        & 2024-02       & \CheckmarkBold                                          & \CheckmarkBold                                         & \textcolor{black}{\textbf{46.3}}                                                                                       & 50.6                                                                                \\
Claude-3-Opus~\cite{claude3}              & Anthropic            & -                                                                                                                                          & 2024-03       &                                                                        & \multicolumn{1}{l}{}                                                  & 82.9                                                                                       & 89.4                                                                                 \\
CodeGemma~\cite{codegemma_2024}          & Google               & 2B, 7B                                                                                                                                     & 2024-04       & \CheckmarkBold                                          & \multicolumn{1}{l}{}                                                  & 44.5                                                                                       & 65.1                                                                                 \\
Code-Qwen~\cite{codeqwen}                 & Qwen Group           & 7B                                                                                                                                         & 2024-04       & \CheckmarkBold                                          & \multicolumn{1}{l}{}                                                  & 45.1                                                                                       & 51.4                                                                                 \\
Llama3~\cite{llama3}                      & Meta                 & 8B, 70B                                                                                                                                    & 2024-04       & \CheckmarkBold                                          & \multicolumn{1}{l}{}                                                  & 81.7                                                                                       & -               \\                                                                 \bottomrule
\end{tabular} 
}} \vspace{-0.4cm}
\end{table*}

\vspace{0.1cm}
\noindent
 \textbf{Target LLM Selection Process.}
 \revised{
Table~\ref{tab:model_selection} presents our analysis of various code-related LLMs in chronological order, till April 2024, when this study was initiated. Models highlighted in light green meet our full open-source criteria, while those marked in light red represent the most suitable candidates based on our selection criteria. After thorough analysis, we selected \textbf{StarCoder}~\cite{starcoder_one} as the research LLM for this study.}

In accordance with our first criterion, we excluded proprietary models such as GPT-3.5~\cite{gpt-3.5-turbo} and GPT-4~\cite{gpt4}, along with recent open-source models like LLaMA3~\cite{llama3}, CodeQwen~\cite{codeqwen}, and DeepSeek-Coder~\cite{deepseekcoder}, which do not fully comply with open-source data requirements. Among the fully open-source options, StarCoder~\cite{starcoder_one} and StarCoder2~\cite{starcoder2} ranked highest on two widely used code generation benchmarks, HumanEval~\cite{chen2021evaluating} and MBPP~\cite{MBPP_1}, thereby fulfilling our second criterion.

When comparing StarCoder~\cite{starcoder_one} with StarCoder2~\cite{starcoder2}, we found that StarCoder’s earlier release positioned it as a foundational model in the field, serving as the basis for several other strong/newer LLMs after fine-tuning, such as PanGu-Coder2~\cite{shen2023pangu}, WizardCoder~\cite{luo2023wizardcoder}, and OctoPack~\cite{muennighoff2023octopack}. This suggests that findings from StarCoder extend to these derivative models. Specifically, if SE benchmark data has leaked to StarCoder, it also affects its descendant LLMs as well, thereby enhancing the broader applicability of our findings and conclusions. Consequently, we opted to select StarCoder over StarCoder2.

Lastly, while some other LLMs do not directly build upon StarCoder, they utilize the pre-training dataset of StarCoder or adopt data curation techniques inspired by StarCoder’s pre-training practices. For instance, CodeShell~\cite{xie2024codeshell} incorporates StarCoder’s pre-training data, while DeepSeek-Coder~\cite{deepseekcoder} employs data-filtering methods akin to those used by StarCoder to gather more recent GitHub data, extending to February 2023. This suggests that findings from StarCoder can be generalized to CodeShell and are likely applicable to DeepSeek-Coder as well, as DeepSeek-Coder follows a similar methodology to extend its datasets. 
In summary, we believe StarCoder is a suitable LLM for us to better understand the SE benchmark leakage status.

\vspace{0.2cm}
\noindent
\textbf{StarCoder's Pre-training Data.} 
StarCoder’s pre-training data is sourced from \textit{The Stack~\cite{stack}} dataset. \textit{The Stack~\cite{stack}} dataset comprises over 6TB of permissively licensed code spanning 358 programming languages. The Stack is collected from public GitHub repositories between 2015 and 2022. For pre-training, StarCoder focused on the 86 programming languages that either contained more than 500MB of data or ranked in the top 50 on popularity indices such as \textit{Githut 2.0}\footnote{\url{https://githut.info/}} or the \textit{December 2022 TIOBE Index}\footnote{\url{https://web.archive.org/web/20221229040526/https://www.tiobe.com/tiobe-index/}}. GitHub repositories for these popular programming languages are recognized as valuable data sources within the SE community, ensuring the representativeness of StarCoder’s pre-training data.

Specifically, StarCoder was pre-trained on a dataset of 305M files, totaling 800+ GB.
In this study, we focus on SE benchmarks in three popular programming languages: Python, Java, and C/C++. Therefore, we restrict our analysis to the corresponding subsets within StarCoder's pre-training data. 
These subsets include 12M files for Python, 20M files for Java, and 14M files for C/C++. 
Despite narrowing our scope to these languages, the data volume remains substantial, supporting our decision to use an efficient automated tool to identify potential duplicate pairs before proceeding with manual labeling.

\vspace{-0.2cm}
 \subsection{Benchmark Data Selection}
 \label{benchmark_selection}

The whole SE community has developed numerous high-quality benchmarks. However, the extensive manual annotation required for each makes it impractical to cover all valuable SE benchmarks. To investigate the leakage status of SE benchmarks in relation to LLMs, we focus on selecting benchmarks that have been previously used for LLM evaluation. Our selection is further constrained to benchmarks within three widely used PLs: Python, Java, and C/C++.

\vspace{0.1cm}
\noindent
\textbf{Studied SE Benchmark Selection.} 
To identify relevant benchmarks, we leverage a recent and comprehensive survey on LLMs for SE tasks~\cite{hou2024large}, which analyzed 395 research papers from January 2017 to January 2024. 
From the papers referenced in this survey, we selected benchmarks and datasets used for evaluating LLMs. We further expanded our selection by examining the citations and related works of these papers, excluding those lacking replication packages or inaccessible datasets.
In building our benchmark collection, we prioritized including datasets across a variety of SE tasks, rather than concentrating on any one task with extensive datasets.

To ensure clarity in our analysis, benchmarks with multiple variants—such as those involving different programming languages or scenarios—were assigned distinguishing tags appended to their original names.
For example, the CodeEditorBench~\cite{CodeEditorBench} benchmark includes variants in three programming languages: Python, Java, and C++. Additionally, it features four scenarios: Code Debug, Code Translate, Code Polish, and Code Requirement Switch. Since each variant contains distinct benchmark data samples, we assign a unique name to each variant.
The naming convention combines the following components:
(1) benchmark original name, (2) scenario name, and (3) programming language name. Name tags for the scenario and programming language are included only if there are multiple variations across different scenarios or programming languages.
For example, the CodeEditorBench~\cite{CodeEditorBench} variant containing Python data samples for the Code Debug scenario is renamed as ``CodeEditorBench-\textit{debug-py}'', following the naming convention. Here, ``\textit{debug}'' indicates the Code Debug scenario, and ``\textit{py}'' represents the Python data.

Through this process, we ultimately compiled a set of 83 SE datasets covering a broad range of SE tasks, including code generation, program repair, code editing, code translation, code review, debugging, code execution, test output prediction, secure code generation, GitHub issue fixing, clone detection, log generation, vulnerability repair, and vulnerability detection.
A brief overview of the benchmarks we studied is provided in Tables~\ref{tab:main_results_py}, ~\ref{tab:main_results_java}, and ~\ref{tab:main_results_c}.

\vspace{-0.2cm}
\subsection{Implementation Details}
We retrieve StarCoder's pre-trained data from its official HuggingFace homepage.\footnote{\url{https://huggingface.co/bigcode/starcoder}} We execute computations on an NVIDIA GeForce A5000 GPU with 24 GB of memory. We acquire the data for each SE benchmark dataset directly from its replication packages or official websites. For the Minhash+LSH method, we adopt the BigCode Team’s implementation~\cite{bigcode}, which leverages the \textit{\textsf{DataSketches}}~\cite{datasketch} library.

\vspace{-0.2cm}
\subsection{Research Questions}
Our work aims to mainly answer four Research Questions (RQs). 
\begin{itemize}[leftmargin=*]
\item \textbf{RQ1: To what extent does data leakage exist in the studied SE benchmarks?} 
In RQ1, we evaluate an extensive set of 83 SE benchmarks to investigate potential data leakage into an advanced LLM StarCoder~\cite{starcoder_one}.
\item \textbf{RQ2: What factors contribute to high leakage rates in SE benchmarks?}
In RQ2, we analyze the top benchmarks with high leakage rates, discussing the potential reasons behind their high leakage ratios. 
\item \textbf{RQ3: How does the leakage of benchmark data affect the effectiveness of LLMs?} 
In RQ3, we measure the effectiveness differences between leaked and non-leaked portions of the data to explore the impact caused by data leakage.
\item \textbf{RQ4: How effective is the automated metric in detecting data leakage when lacking access to LLM pre-training data?} 
Since pre-training data for many LLMs is inaccessible, this RQ investigates whether SE benchmark leakage can be inferred solely from LLM behaviors (i.e., without access to pre-training data), such as the Perplexity scores of LLMs. 
\end{itemize}

\section{Experimental Results}\label{section5}

In this section, we first present the overall results of the data leakage detection analysis. Following that, we provide detailed results and answers to each research question (RQ).

\vspace{0.2cm}
\noindent
\textbf{Overall Results.}
Before diving into the specific RQs, we first present an overview of the experimental results. The selected LLM's pre-training datasets consist of 12M samples for Python, 20M samples for Java, and 14M samples for C/C++. In comparison, the diverse SE benchmarks we studied collectively comprise 46k samples for Python, 42k samples for Java, and 21k samples for C/C++.
To investigate potential data leakage, each SE benchmark sample was compared against all pre-training data samples for its corresponding programming language. This process resulted in an astounding total of over 1.7 trillion comparisons. The sheer scale of this computational effort highlights the complexity and resource-intensive nature of studying data leakage regarding LLMs.

\begin{figure}[t] 
    \centering
    \includegraphics[width=12cm]{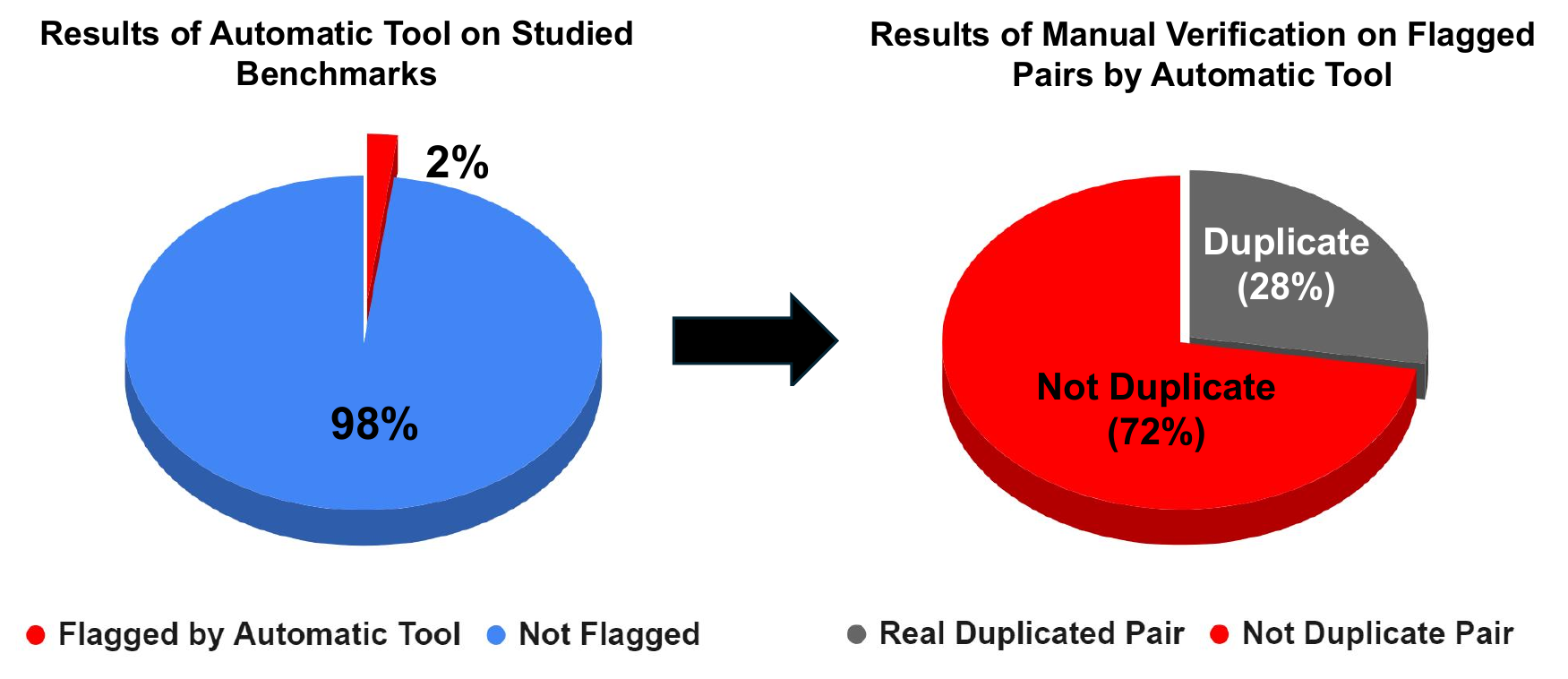} 
    \vspace{-0.2cm}
    \caption{Data Leakage Overall Results} 
    \vspace{-0.3cm}
    \label{fig:data_review}
\end{figure}

From an overall perspective, as depicted in Figure~\ref{fig:data_review}, only 2\% of the benchmark samples from all the SE benchmarks studied were flagged by the automated tool MinHash+LSH as potentially forming at least one duplicate pair with the pre-training data of StarCoder. Moreover, of the pairs flagged by MinHash+LSH, 28\% were confirmed as duplicates after manual labeling, while the remaining 72\% were determined not to be duplicates.

Next, we will discuss the detailed results and answers to each RQ.

\subsection{RQ1: Data Leakage Status in SE Benchmarks}

\begin{table}[t]
\centering
\caption{Results of Data Leakage in SE Benchmarks (Python). \#Auto represents the number of potential duplicate pairs identified by MinHash+LSH, while \#Manual denotes the number of true duplicate pairs labeled by developers.}
\label{tab:main_results_py}
\scalebox{0.8}{
\rotatebox{0}{
\begin{tabular}{llrrrrr}
\hline
\textbf{Benchmark} & \textbf{Task(s)} & \textbf{Size} & \multicolumn{1}{c}{\textbf{\begin{tabular}[c]{@{}c@{}}\#Auto\end{tabular}}} & \multicolumn{1}{c}{\textbf{\begin{tabular}[c]{@{}c@{}}\#Manual\end{tabular}}} & \multicolumn{1}{c}{\textbf{\begin{tabular}[c]{@{}c@{}}Leaked\\ Count\end{tabular}}} & \multicolumn{1}{c}{\textbf{\begin{tabular}[c]{@{}c@{}}Leaked\\ Ratio\end{tabular}}} \\ \hline
APPS~\cite{hendrycks2021measuring} & Code Generation & 10k & 240 & 193 & 108 & \cellcolor{red!21.6} 10.8\% \\
BigCodeBench\textit{-py}~\cite{zhuo2024bigcodebench} & Code Generation & 1.14k & 0 & 0 & 0 & 0\% \\
BioCoder~\cite{BioCoder} & Code Generation & 207 & 1 & 0 & 0 & 0\% \\
BugsInPy~\cite{BugsInPy} & Program Repair & 501 & 322 & 56 & 55 & \cellcolor{red!22} 11.0\% \\
CanItEdit~\cite{CanItEdit} & Code Editing & 105 & 0 & 0 & 0 & 0\% \\
ClassEval~\cite{ClassEval} & Code Generation & 100 & 0 & 0 & 0 & 0\% \\
CodeBenchGen~\cite{CodeBenchGen} & Code Generation & 1.93k & 0 & 0 & 0 & 0\% \\
CodeEditorBench\textit{-debug-py}~\cite{CodeEditorBench}  & Code Editing & 356 & 93 & 92 & 38 & \cellcolor{red!21.4} 10.7\% \\
CodeEditorBench\textit{-polish-py}~\cite{CodeEditorBench} & Code Editing & 413 & 1 & 1 & 1 & \cellcolor{red!0.4} 0.2\% \\
CodeEditorBench\textit{-switch-py}~\cite{CodeEditorBench}  & Code Editing & 488 & 115 & 103 & 35 & \cellcolor{red!14.4} 7.2\% \\
CodeEditorBench\textit{-translate-py}~\cite{CodeEditorBench}  & Code Translation & 709 & 0 & 0 & 0 & 0\% \\
CodeReview\textit{-py}~\cite{CodeReview} & Code Review & 2.9k & 0 & 0 & 0 & 0\% \\
CodeReviewNew\textit{-py}~\cite{CodeReviewNew} & Code Review & 2.15k & 3 & 1 & 1 & \cellcolor{red!0.1} 0.05\% \\
CodeScope\textit{-py}~\cite{Codescope} & Code Generation & 400 & 0 & 0 & 0 & 0\% \\
CoNala\textit{-curated}~\cite{CoNala} & Code Generation & 2.88k & 1 & 1 & 1 & \cellcolor{red!0.06} 0.03\% \\
ConDefects~\cite{Condefects} & Program Repair & 2.86k & 205 & 8 & 8 & \cellcolor{red!0.6} 0.3\% \\
DebugBench\textit{-py}~\cite{DebugBench} & Debugging & 4.25k & 2 & 1 & 1 & \cellcolor{red!0.04} 0.02\% \\
DS-1000~\cite{DS-1000} & Code Generation & 1k & 0 & 0 & 0 & 0\% \\
EvoCodeBench~\cite{EvoCodeBench} & Code Generation & 275 & 21 & 18 & 18 & \cellcolor{red!13} 6.5\% \\
G-TransEval\textit{-py}~\cite{G-TransEval} & Code Translation & 400 & 0 & 0 & 0 & 0\% \\
HumanEval~\cite{codex} & Code Generation & 164 & 3 & 3 & 3 & \cellcolor{red!3.6} 1.8\% \\
LiveCodeBench\textit{-code-generation}~\cite{LiveCodeBench}  & Code Generation & 511 & 0 & 0 & 0 & 0\% \\
LiveCodeBench\textit{-execution}~\cite{LiveCodeBench}  & Code Execution & 479 & 0 & 0 & 0 & 0\% \\
LiveCodeBench\textit{-test-generation}~\cite{LiveCodeBench} & \begin{tabular}[c]{@{}l@{}}Test Output \\ Prediction\end{tabular} & 442 & 0 & 0 & 0 & 0\% \\
MBPP~\cite{MBPP_1} & Code Generation & 974 & 4 & 4 & 4 & \cellcolor{red!0.8} 0.4\% \\
Mconala\textit{-es}~\cite{MCoNaLa} & Code Generation & 341 & 0 & 0 & 0 & 0\% \\
Mconala\textit{-ja}~\cite{MCoNaLa} & Code Generation & 210 & 1 & 0 & 0 & 0\% \\
Mconala\textit{-ru}~\cite{MCoNaLa} & Code Generation & 345 & 30 & 1 & 1 & \cellcolor{red!0.6} 0.3\% \\
Mercury~\cite{Mercury} & Code Generation & 1.89k & 12 & 12 & 10 & \cellcolor{red!1} 0.5\% \\
PythonSaga~\cite{PythonSaga} & Code Generation & 185 & 0 & 0 & 0 & 0\% \\
QuixBugs~\cite{QuixBugs} & Program Repair & 40 & 84 & 84 & 40 & \cellcolor{red!100} \textbf{100.0\%} \\
Refactory~\cite{Refactory} & Program Repair & 4.39k & 198 & 7 & 7 & \cellcolor{red!0.4} 0.2\% \\
SecurityEval~\cite{SecurityEval} & Secure Code Gene. & 121 & 2 & 2 & 2 & \cellcolor{red!3.4} 1.7\% \\
SVEN\textit{-py}~\cite{SVEN} & Secure Code Gene. & 28 & 0 & 0 & 0 & 0\% \\
SWE-Bench~\cite{SWE-bench} & Issue Fix & 2.52k & 2175 & 221 & 220 & \cellcolor{red!17.4} 8.7\% \\
SWE-Bench\textit{-verified}~\cite{SWE-bench-verified}  & Issue Fix & 500 & 59 & 53 & 53 & \cellcolor{red!21.2} 10.6\% \\ \hline
\textbf{Average} & - & - & - & - & - & 4.8\% \\
\bottomrule
\end{tabular}
}} \vspace{-0.4cm}
\end{table}

\begin{table}[t]
\centering
\caption{Results of Data Leakage in SE Benchmarks (Java). \#Auto represents the number of potential duplicate pairs identified by MinHash+LSH, while \#Manual denotes the number of true duplicate pairs labeled by developers.}
\label{tab:main_results_java}
\scalebox{0.8}{
\rotatebox{0}{
\begin{tabular}{llrrrrr}
\hline
\textbf{Benchmark} & \textbf{Task(s)} & \textbf{Size} & \multicolumn{1}{c}{\textbf{\begin{tabular}[c]{@{}c@{}}\#Auto\end{tabular}}} & \multicolumn{1}{c}{\textbf{\begin{tabular}[c]{@{}c@{}}\#Manual\end{tabular}}} & \multicolumn{1}{c}{\textbf{\begin{tabular}[c]{@{}c@{}}Leaked\\ Count\end{tabular}}} & \multicolumn{1}{c}{\textbf{\begin{tabular}[c]{@{}c@{}}Leaked\\ Ratio\end{tabular}}} \\ \hline
AixBench\textit{-auto}~\cite{AixBench} & Code Generation & 175 & 0 & 0 & 0 & 0\% \\
AixBench\textit{-manual}~\cite{AixBench} & Code Generation & 161 & 0 & 0 & 0 & 0\% \\
API-Misuse-Repair\textit{-complete}~\cite{API-Misuse-Repair}  & Program Repair & 11.8k & 21 & 15 & 12 & \cellcolor{red!0.2} 0.1\% \\
API-Misuse-Repai\textit{-single}~\cite{API-Misuse-Repair} & Program Repair & 5.45k & 29 & 7 & 6 & \cellcolor{red!0.2} 0.1\% \\
BigCloneBench~\cite{BigCloneBench} & Clone Detection & 912 & 667 & 513 & 508 & \cellcolor{red!75.7} \textbf{55.7\%} \\
BigCodeBench-java~\cite{zhuo2024bigcodebench} & Code Generation & 1.14k & 0 & 0 & 0 & 0\% \\
Bugs.Jar~\cite{Bugs-Jar} & Program Repair & 1.16k & 1 & 1 & 1 & \cellcolor{red!0.2} 0.09\% \\
CodeEditorBench\textit{-debug-java}~\cite{CodeEditorBench}  & Code Editing & 246 & 2 & 2 & 2 & \cellcolor{red!1.6} 0.8\% \\
CodeEditorBench\textit{-polish-java}~\cite{CodeEditorBench}   & Code Editing & 279 & 3 & 3 & 3 & \cellcolor{red!2.2} 1.1\% \\
CodeEditorBench\textit{-switch-java}~\cite{CodeEditorBench} & Code Editing & 433 & 117 & 113 & 43 & \cellcolor{red!20} 9.9\% \\
CodeEditorBench\textit{-translate-java}~\cite{CodeEditorBench}  & Code Editing & 657 & 0 & 0 & 0 & 0\% \\
CodeReview\textit{-java}~\cite{CodeReview} & Code Review & 2.21k & 1330 & 0 & 0 & 0\% \\
CodeReviewNew\textit{-java}~\cite{CodeReviewNew}  & Code Review & 1.21k & 457 & 0 & 0 & 0\% \\
CodeScope\textit{-java}~\cite{Codescope} & Code Generation & 400 & 0 & 0 & 0 & 0\% \\
Concode~\cite{Concode} & Code Generation & 2k & 0 & 0 & 0 & 0\% \\
DebugBench\textit{-java}~\cite{DebugBench} & Debugging & 4.25k & 6 & 6 & 6 & \cellcolor{red!0.28} 0.14\% \\
Defects4J-2.0~\cite{Defects4J} & Program Repair & 483 & 2 & 2 & 2 & \cellcolor{red!0.82} 0.41\% \\
G-TransEval\textit{-java}~\cite{G-TransEval} & Code Translation & 400 & 1 & 0 & 0 & 0\% \\
GitBug~\cite{GitBug} & Program Repair & 199 & 0 & 0 & 0 & 0\% \\
Humaneval-x\textit{-java}~\cite{zheng2023codegeex} & Code Generation & 164 & 0 & 0 & 0 & 0\% \\
LogBench-o~\cite{LogBench} & Log Generation & 3.86k & 160 & 142 & 142 & \cellcolor{red!7.4} 3.7\% \\
LogBench-t~\cite{LogBench} & Log Generation & 3.84k & 103 & 92 & 92 & \cellcolor{red!4.8} 2.4\% \\
MultiPLE\textit{-humaneval-java}~\cite{MultiPLE}   & Code Generation & 158 & 0 & 0 & 0 & 0\% \\
MultiPLE\textit{-MBPP-java}~\cite{MultiPLE} & Code Generation & 386 & 0 & 0 & 0 & 0\% \\
Robustness-Copilot~\cite{Robustness-Copilot} & Code Generation & 892 & 0 & 0 & 0 & 0\% \\
VJBench~\cite{VJBench} & Vulnerability Repair & 15 & 0 & 0 & 0 & 0\% \\
Vul4J~\cite{Vul4J} & Vulnerability Repair & 35 & 0 & 0 & 0 & 0\% \\ \hline
\textbf{Average} & - & - & - & - & - & 2.8\% \\
\bottomrule
\end{tabular}
}} \vspace{-0.4cm}
\end{table}

\begin{table}[t]
\centering
\caption{Results of Data Leakage in SE Benchmarks (C/C++). \#Auto represents the number of potential duplicate pairs identified by MinHash+LSH, while \#Manual denotes the number of true duplicate pairs labeled by developers.}
\label{tab:main_results_c}
\scalebox{0.85}{
\rotatebox{0}{
\begin{tabular}{llrrrrr}
\hline
\textbf{Benchmark} & \textbf{Task(s)} & \textbf{Size} & \multicolumn{1}{c}{\textbf{\begin{tabular}[c]{@{}c@{}}\#Auto\end{tabular}}} & \multicolumn{1}{c}{\textbf{\begin{tabular}[c]{@{}c@{}}\#Manual\end{tabular}}} & \multicolumn{1}{c}{\textbf{\begin{tabular}[c]{@{}c@{}}Leaked\\ Count\end{tabular}}} & \multicolumn{1}{c}{\textbf{\begin{tabular}[c]{@{}c@{}}Leaked\\ Ratio\end{tabular}}} \\ \hline
BigVul-CVEFix\textit{-Repair}~\cite{BigVul-CVEFix-Repair}  & Vulnerability Repair & 1.93k & 21 & 14 & 14 & \cellcolor{red!1.4} 0.7\% \\
BigCodeBench-c~\cite{zhuo2024bigcodebench} & Code Generation & 1.14k & 0 & 0 & 0 & 0\% \\
CodeEditorBench\textit{-debug-c}~\cite{CodeEditorBench}  & Code Editing & 336 & 28 & 27 & 15 & \cellcolor{red!9} 4.5\% \\
CodeEditorBench\textit{-polish-c}~\cite{CodeEditorBench} & Code Editing & 309 & 1 & 1 & 1 & \cellcolor{red!0.6} 0.3\% \\
CodeEditorBench\textit{-switch-c}~\cite{CodeEditorBench} & Code Editing & 530 & 90 & 90 & 44 & \cellcolor{red!16.6} \textbf{8.3\%} \\
CodeEditorBench\textit{-translate-c}~\cite{CodeEditorBench} & Code Editing & 702 & 1 & 1 & 1 & \cellcolor{red!0.2} 0.1\% \\
CodeReview\textit{-c}~\cite{CodeReview} & Code Review & 1.8k & 0 & 0 & 0 & 0\% \\
CodeReviewNew\textit{-c}~\cite{CodeReviewNew} & Code Review & 1.05k & 0 & 0 & 0 & 0\% \\
CodeScope\textit{-c}~\cite{Codescope} & Code Generation & 400 & 0 & 0 & 0 & 0\% \\
D2A~\cite{D2A} & Vulnerability Detection & 4.6k & 8 & 4 & 4 & \cellcolor{red!0.2} 0.1\% \\
DebugBench\textit{-c}~\cite{DebugBench} & Debugging & 4.25k & 25 & 25 & 22 & \cellcolor{red!1} 0.5\% \\
Devign~\cite{Devign} & Vulnerability Detection & 2.73k & 16 & 7 & 7 & \cellcolor{red!0.6} 0.3\% \\
G-TransEval\textit{-c}~\cite{G-TransEval} & Code Translation & 400 & 6 & 0 & 0 & 0\% \\
Humaneval-x\textit{-c}~\cite{zheng2023codegeex} & Code Generation & 164 & 0 & 0 & 0 & 0\% \\
MultiPLE\textit{-humaneval-c}~\cite{MultiPLE} & Code Generation & 161 & 0 & 0 & 0 & 0\% \\
MultiPLE\textit{-mbpp-c}~\cite{MultiPLE} & Code Generation & 397 & 0 & 0 & 0 & 0\% \\
Pareval\textit{-generation}~\cite{Pareval} & Code Generation & 420 & 0 & 0 & 0 & 0\% \\
Pareval\textit{-translation}~\cite{Pareval} & Code Generation & 143 & 0 & 0 & 0 & 0\% \\
SVEN\textit{-c}~\cite{SVEN} & Secure Code Gene. & 18 & 0 & 0 & 0 & 0\% \\
VulnPatchPairs~\cite{VulnPatchPairs} & Vulnerability Detection & 100 & 0 & 0 & 0 & 0\% \\ \hline
\textbf{Average} & - & - & - & - & - & 0.7\% \\
\bottomrule
\end{tabular}
}} \vspace{-0.4cm}
\end{table}

Table~\ref{tab:main_results_py} to Table~\ref{tab:main_results_c} present the data leakage status of the studied SE benchmarks for Python, Java, and C/C++, respectively. The column ``\textit{\#Auto}'' represents the number of potentially duplicate pairs identified by the automated tool MinHash+LSH. The column ``\textit{\#Manual}'' shows the number of actual duplicate pairs verified through manual labeling. The column ``\textit{Leaked Count}'' refers to the Leakage Count, which is the number of actual leaked SE benchmark samples. Lastly, the column ``\textit{Leaked Ratio}'' refers to the Leakage Ratio, which shows the proportion of benchmark data identified as leaked to an LLM.
Both the Leakage Count and Leakage Ratio serve as key metrics for assessing the extent of data leakage in the benchmarks. A higher value for these metrics indicates a more severe data leakage problem in the given SE benchmark.
It is important to note that a single SE benchmark sample may appear multiple times in an LLM's pre-training dataset. For instance, an identical function may be present in different software repositories used in LLMs's pre-training. These occurrences can result in multiple duplicate pairs being identified, even though they correspond to a single leaked SE benchmark sample. As a result, we may observe that the Leakage Count is occasionally smaller than the number of duplicate pairs identified during manual labeling.

 \vspace{-0.1cm}
\subsubsection{Data Leakage in Python Benchmarks}
\revised{
Table~\ref{tab:main_results_py} reveals that data leakage in Python benchmarks is generally minimal, with an average leakage ratio of only 4.8\%. Our results indicate that \textbf{many Python benchmarks have low or negligible leakage rates}, making them suitable for reliable evaluation of LLMs. As shown in Table~\ref{tab:main_results_py}, several benchmarks, such as BigCodeBench\textit{-py}, BioCoder, CanItEdit, ClassEval, CodeBenchGen, CodeEditorBench\textit{-translate-py}, CodeReview\textit{-py}, CodeReviewNew\textit{-py}, CodeScope\textit{-py}, DS-1000, G-TransEval\textit{-py}, LiveCodeBench series, Mconala\textit{-es}, Mconala\textit{-ja}, PythonSaga, and SVEN\textit{-py}, exhibit no evidence of the leakage issue (i.e., Leakage Ratio=0\%) in our study, highlighting that the threat of data leakage is relatively small when using those benchmarks to evaluate LLMs.}

\revised{
\textbf{However, a few Python benchmarks demonstrate much higher leakage rates, raising concerns about their bias in evaluation.} Notable benchmarks include 
APPS (10.8\%), BugsInPy (11.0\%), CodeEditorBench\textit{-debug-py} (10.7\%), CodeEditorBench\textit{-switch-py} (7.2\%),
EvoCodeBench (6.5\%),
QuixBugs (100\%), SWE-Bench (8.7\%), and SWE-Bench\textit{-verified} (10.6\%). These elevated leakage ratios suggest that a relatively large portion of the data samples in these benchmarks was encountered during LLM pre-training, which could compromise the fairness and validity of evaluation results. 
This underscores the importance of removing identified duplicate benchmark data in future work to ensure unbiased evaluation when utilizing these benchmarks.
}

 \vspace{-0.1cm}
\subsubsection{Data Leakage in Java Benchmarks}
Data leakage in Java benchmarks is also generally low, with an average leakage ratio of just 2.8\%, as shown in Table~\ref{tab:main_results_java}. This is notably lower than the average leakage ratio observed in Python benchmarks. The results reveal that \textbf{many Java benchmarks exhibit low or negligible leakage rates}, making them suitable for reliable evaluation of LLMs. Many Java benchmarks, including AixBench\textit{-auto}, AixBench\textit{-manual}, and several others, show no evidence of leakage issues (i.e., Leakage Ratio=0\%). This finding highlights that the risk of data leakage is relatively small when using these benchmarks.

\textbf{However, two Java benchmarks exhibit much higher leakage rates.
}
BigCloneBench has a leakage ratio of 55.7\%, while CodeEditorBench\textit{-switch-java} has a leakage ratio of 9.9\%. 
Thus, it is important to remove those identified leaked benchmark samples before using these two benchmarks in future research related to LLMs.

 \vspace{-0.1cm}
\subsubsection{Data Leakage in C/C++ Benchmarks}
\revised{
Data leakage in C/C++ benchmarks is minimal, with an average leakage ratio of only 0.7\%, as shown in Table~\ref{tab:main_results_c}. This represents the lowest average leakage ratio among the Python, Java, and C/C++ benchmarks analyzed. The results indicate that \textbf{most C/C++ benchmarks exhibit low or negligible leakage rates}, making them suitable for reliable evaluation of LLMs.
\textbf{However, two C/C++ benchmarks demonstrate relatively higher leakage rates}: CodeEditorBench\textit{-debug-c} (4.5\%) and CodeEditorBench\textit{-switch-c} (8.3\%). These elevated leakage rates emphasize the importance of addressing data leakage by removing identified duplicate benchmark data when using these two benchmarks. In contrast, the threat of data leakage for other C/C++ benchmarks remains significantly lower.
}

 \vspace{-0.1cm}
\subsubsection{Data Leakage Status in Specific SE Tasks}

\revised{
We will now discuss the leakage rates across different tasks. We first focus on Code Generation and Program Repair, as these are both highly popular tasks and constitute the majority of the benchmarks analyzed.
For Code Generation, out of 33 benchmarks, 26 have a leakage rate of 0\%, and the average leakage rate across all benchmarks is just 0.62\%. This indicates that the issue of data leakage in the Code Generation benchmarks we studied is relatively minor.
In contrast, for Program Repair, only 1 out of 9 benchmarks has a leakage rate of 0\%, while the average leakage rate reaches 12.5\%. This suggests that data leakage poses a significant challenge for Program Repair benchmarks.}

\revised{
Furthermore, when examining the top 10 benchmarks with the highest leakage rates, the task distribution is as follows: Code Editing (4 benchmarks), Program Repair (2 benchmarks), Issue Fix (2 benchmarks), Clone Detection (1 benchmark), and Code Generation (1 benchmark). 
This distribution indicates that Program Repair, Code Editing, and Issue Fix tasks are relatively more vulnerable to higher leakage rates, as they contain more benchmarks with top leakage compared to other tasks.
}

 \vspace{-0.1cm}
\subsubsection{LLM Evaluation with LessLeak-Bench}
Our experimental results indicate that while the average leakage rate in SE benchmarks is low, certain benchmarks have been significantly impacted by data leakage. To support researchers, we have removed the leaked samples and developed LessLeak-Bench, a cleaned version of SE benchmarks. We recommend using LessLeak-Bench in future studies to ensure more reliable LLM evaluations, as many leaked benchmark samples are no longer suitable for this purpose.

\vspace{0.3cm}
\noindent
\begin{tcolorbox} [boxrule=0.8pt,
                top=0.2pt,
                  bottom=0.2pt]
    \textbf{Answer to RQ1}: 
    \revised{
   In general, data leakage in SE benchmarks is minimal, with average leakage ratios of only 4.8\%, 2.8\%, and 0.7\% for Python, Java, and C/C++ benchmarks, respectively. However, some benchmarks exhibit relatively higher leakage ratios. For instance, QuixBugs and BigCloneBench have leakage ratios of 100.0\% and 55.7\%, respectively.
   Additionally, the Program Repair task generally shows higher leakage ratios, with an average leakage rate of 12.5\%. Researchers should exercise caution when applying LLMs to program repair tasks, as these benchmarks may require the removal of duplicate data samples to ensure unbiased evaluations.}
\end{tcolorbox}

\subsection{RQ2: Understanding High Data Leakage in SE Benchmarks}

\subsubsection{Which Types of Software Repositories Are Likely Contributing to Data Leakage?}

\revised{
We conducted an investigation into which software repositories in the pre-training data of the LLM contribute most significantly to data leakage in SE benchmarks. To do this, we grouped all identified duplicate pairs, where the SE benchmark samples are the leaked benchmark data, and the pre-training data in these pairs indicate the source repositories in pre-training data. By analyzing the characteristics of these repositories, we can offer actionable recommendations. 
}

\revised{
Table~\ref{tab:pair_in_repos} highlights the software repositories associated with the most duplicate pairs. One key finding is that 92 duplicate pairs originated from the inclusion of the ``PatrickShaw/QuixBugs'' repository. This accounts for the high data leakage ratio observed in the QuixBugs benchmark, as the pre-training data seems to contain a cloned version of the QuixBugs repository. Additionally, we observed a high frequency of algorithm-related keywords, such as ``leetcode,'' ``programming," and ``data-structures," in repository names. For instance, ``leetcode'' appears 8 times in the top 20 repositories. 
}

\revised{
Based on these observations, we offer several actionable recommendations. For LLM developers, it is crucial to scan repository names during the preparation of pre-training data and filter out repositories that include common SE benchmark names, such as ``PatrickShaw/QuixBugs'' listed in Table~\ref{tab:pair_in_repos}. This will help prevent the inadvertent inclusion of benchmarks through cloned repositories, thereby introducing benchmark data into the pre-training data. Additionally, repositories containing common coding website names, such as 'leetcode,' or algorithm-related keywords like "data-structures," should be excluded. Many benchmarks are sourced from these platforms or algorithm-related repositories, which increases the likelihood of data leakage.
For LLM users and researchers evaluating LLMs, we recommend avoiding the direct use of benchmark data primarily sourced from coding websites like LeetCode or from algorithm-related tutorial repositories. This approach will help reduce the risk of data leakage during evaluations.
}

\begin{table}[t]
\centering
\caption{Software repositories used in LLM pre-training that have a high number of identified duplicate pairs after manual labeling.}
\label{tab:pair_in_repos}
\scalebox{0.9}{
\rotatebox{0}{
\begin{tabular}{lr}
\hline
\textbf{Project Name} & \textbf{\begin{tabular}[c]{@{}r@{}}\#Manual Labeled\\ Duplicate Pair\end{tabular}} \\ \hline
cragkhit/elasticsearch & 509 \\
sgholamian/log-aware-clone-detection & 229 \\
PatrickShaw/QuixBugs & 92 \\
devangi2000/Data-Structures-Algorithms-Handbook & 37 \\
RafaelHuang87/Leet-Code-Practice & 20 \\
khushi-411/LeetCode & 17 \\
sugia/leetcode & 16 \\
kppw99/enVAS & 15 \\
naddym/competitive-programming & 14 \\
VinceW0/Leetcode\_Python\_solutions & 13 \\
AvadheshChamola/LeetCode & 13 \\
NikolayVaklinov10/Python\_Challenges & 10 \\
tirthbharatiya/interview\_questions & 10 \\
wingkwong/competitive-programming & 9 \\
abdzitter/Daily-Coding-DS-ALGO-Practice & 9 \\
vedantc6/LCode & 9 \\
Taewan-P/LeetCode\_Repository & 9 \\
apoorvkk/LeetCodeSolutions & 9 \\
jen-sjen/data-structures-basics-leetcode & 9 \\
chaosWsF/Python-Practice   & 9 \\
\hline
\end{tabular}
}
} \vspace{-0.4cm}
\end{table}

\subsubsection{Why Do Certain SE Benchmarks Exhibit High Data Leakage?}

This subsection explores benchmarks with high data leakage rates, aiming to investigate the potential underlying causes. While most benchmarks exhibit minimal data leakage, this research question focuses specifically on the top 10 benchmarks with the highest leakage. By identifying the reasons and characteristics behind their high leakage rates, we aim to equip researchers with insights to assess whether their chosen benchmarks may face similar issues.

It is important to clarify that the issue does not stem from the benchmarks themselves. Many of these benchmarks were developed and published prior to the advent of LLMs. As a result, the high leakage rates are not due to any flaws in the benchmarks, but rather an unintended consequence of the overlap between the historical data used to create the benchmarks and the data included in LLM training.

Table~\ref{tab:causes} provides an overview of the top 10 benchmarks with the highest leakage rates, detailing their data sources, basic descriptions, and the potential causes behind their high leakage. Specifically, QuixBugs exhibits significant leakage due to the inclusion of the ``PatrickShaw/QuixBugs'' repository in the pre-training dataset. BigCloneBench's high leakage is attributed to the overlap between its source data and the ``cragkhit/elasticsearch'' repository included in the pre-training data. This overlap likely results from BigCloneBench’s reliance on IJaDataset 2.0~\cite{IJaDataset}, which contains the Elasticsearch project. A similar issue affects BugsInPy, where six source projects used in its dataset construction were also identified in LLM's pre-training data.
For other benchmarks, the primary cause of high data leakage lies in their dependence on widely used coding platforms. Benchmarks such as APPS, CodeEditorBench\textit{-switch-java}, CodeEditorBench\textit{-debug-py}, CodeEditorBench\textit{-switch-c}, and CodeEditorBench\textit{-switch-py} are mainly derived from LeetCode. Since data from LeetCode is included in LLM pre-training datasets, these benchmarks inherently exhibit relatively high leakage rates. Finally, SWE-Bench and SWE-Bench\textit{-verified} also experience substantial overlap because they use GitHub issues as their primary data source, many of which are already incorporated into LLM pre-training datasets.

These findings highlight four primary causes of high data leakage: 1) the direct inclusion of benchmark data in the pre-training dataset (e.g., QuixBugs), 2) overlap with the repositories used to create benchmarks (e.g., BigCloneBench and BugsInPy), 3) dependence on coding platforms like LeetCode (e.g., APPS and CodeEditorBench variants), or 4) shared use of data sources such as GitHub issues (e.g., SWE-Bench).

To address these challenges, 
careful selection of benchmarks is essential for LLM users. They should be cautious about using benchmarks derived from popular coding websites or repositories with known overlaps, and follow the approach in this paper to compare their chosen benchmarks against pre-training datasets to assess the risk of benchmark leakage. 
Doing so helps ensure more reliable evaluations while minimizing the impact of data leakage on benchmark results.

\begin{table*}[t]
\centering
\caption{Potential Causes of High Data Leakage in the Top-10 Leaked SE Benchmarks.}
\label{tab:causes}
\scalebox{0.7}{
\begin{tabular}{lc|l}
\midrule
\textbf{\begin{tabular}[c]{@{}l@{}}Relatively\\ High Leaked\\ Benchmarks\end{tabular}} &  \begin{tabular}[c]{@{}c@{}}\textbf{Data}\\ \textbf{Sources}\end{tabular}  & \multicolumn{1}{c}{\textbf{Benchmark Description \& Potential Reason}} \\ \midrule

QuixBugs~\cite{QuixBugs} & \begin{tabular}[c]{@{}c@{}}Coding\\Challenge\end{tabular}  & \begin{tabular}[c]{@{}l@{}} \textbf{Description:} QuixBugs includes implementations of classic algorithms with a bug on a single line,\\ and the corresponding fixes, sourced from the Quixey Challenge. 
\\ \textbf{Reason:} One copy of QuixBugs on GitHub is directly included in the LLM pre-training data.  \end{tabular} \\ \midrule
BigCloneBench~\cite{BigCloneBench} &  \begin{tabular}[c]{@{}c@{}}IJaDataset\\2.0\end{tabular} & \begin{tabular}[c]{@{}l@{}}
\textbf{Description:} IJaDataset 2.0~\cite{IJaDataset} consists of 25,000 open-source Java systems spanning 3 million\\ files and 250 million LOC. \\ 
\textbf{Reason:} BigCloneBench includes 504 leaked samples sourced from the project \\``cragkhit/elasticsearch'' in the LLM pre-training data. It is likely that \\ the elasticsearch project is included in IJaDataset 2.0. We are unable to download the \\ original IJaDataset 2.0 to confirm this explanation because the link~\cite{IJaDataset} of IJaDataset 2.0 was expired.\end{tabular} \\ \midrule

BugsInPy~\cite{BugsInPy}  & \begin{tabular}[c]{@{}c@{}}GitHub\\Projects\end{tabular} & \begin{tabular}[c]{@{}l@{}}
\textbf{Description:} BugsInPy includes 493 Python bugs collected from 17 different Python projects.\\ 
\textbf{Reason:} 
Some source projects in BugsInPy are included in the pre-training data of LLMs. Specifically,\\ we identified six source projects from BugsInPy that are present in LLMs' pre-training datasets. \end{tabular} \\ \midrule
APPS~\cite{hendrycks2021measuring} &  \begin{tabular}[c]{@{}c@{}}Coding\\Websites\end{tabular} 
 & \begin{tabular}[c]{@{}l@{}}
\textbf{Description:} APPS contains programming problems sourced from coding websites (e.g., Codeforces).\\
\textbf{Reason:} APPS has been leaked, as the LLM pre-training data includes numerous repositories from\\ LeetCode, which contains many programming problems. By examining the sources of pre-training\\ data that duplicate APPS's data, we identified 36 repositories whose names contain the string ``leetcode".\end{tabular} \\ \midrule
 \begin{tabular}[c]{@{}l@{}}CodeEditorBench\\ \textit{-debug-py}~\cite{CodeEditorBench}\end{tabular}  & \begin{tabular}[c]{@{}c@{}}Coding\\Challenge\end{tabular} & \begin{tabular}[c]{@{}l@{}}
\textbf{Description:} It is a curated set of coding challenges from various sources, including LeetCode. \\ \textbf{Reason:} This benchmark is leaked as the LLM pre-training data includes many repositories from\\ LeetCode. By examining the sources of pre-training data that duplicate this benchmark's data,\\ we identified 9 repositories whose names contain the string ``leetcode".\end{tabular} \\ \midrule
\begin{tabular}[c]{@{}l@{}}SWE-Bench\\ \textit{-verified}~\cite{SWE-bench-verified}\end{tabular}
 & \begin{tabular}[c]{@{}c@{}}GitHub\\Issues\end{tabular}  & \begin{tabular}[c]{@{}l@{}}
\textbf{Description:} SWE-Bench-verified is collected from GitHub issues. It mainly includes the issues, \\the versions prior to their fixes, and the corresponding fixed versions. This benchmark is a \\human-validated subset of SWE-Bench.
\\ \textbf{Reason:} 
GitHub issues are included in the LLM's pre-training data, resulting in some overlap\\ with this benchmark's data.\end{tabular} \\ \midrule

\begin{tabular}[c]{@{}l@{}}CodeEditorBench\\ \textit{-switch-java}\end{tabular}~\cite{CodeEditorBench}    & \begin{tabular}[c]{@{}c@{}}Coding\\Challenge\end{tabular} & \begin{tabular}[c]{@{}l@{}}
\textbf{Description:} It is a curated set of coding challenges from various sources, including LeetCode. \\ 
\textbf{Reason:} This benchmark is leaked as the LLM pre-training data includes many repositories from\\ LeetCode. By examining the sources of pre-training data that duplicate this benchmark's data,\\ we identified 13 repositories whose names contain the string ``leetcode".\end{tabular} \\ \midrule
SWE-Bench~\cite{SWE-bench} & \begin{tabular}[c]{@{}c@{}}GitHub\\Issues\end{tabular} & \begin{tabular}[c]{@{}l@{}}
\textbf{Description:} SWE-Bench is collected from GitHub issues. It mainly includes the issues, \\the versions prior to their fixes, and the corresponding fixed versions. 
\\ \textbf{Reason:} 
GitHub issues are included in the LLM's pre-training data, resulting in some overlap\\ with this benchmark's data.\end{tabular} \\ 
\midrule

\begin{tabular}[c]{@{}l@{}}CodeEditorBench\\ \textit{-switch-c}\end{tabular}~\cite{CodeEditorBench}   & \begin{tabular}[c]{@{}c@{}}Coding\\Challenge\end{tabular} & \begin{tabular}[c]{@{}l@{}}
\textbf{Description:} It is a curated set of coding challenges from various sources, including LeetCode. \\ 
\textbf{Reason:} This benchmark is leaked as the LLM pre-training data includes many repositories from\\ LeetCode. By examining the sources of pre-training data that duplicate this benchmark's data,\\ we identified 10 repositories whose names contain the string ``leetcode".\end{tabular} \\ \midrule
\begin{tabular}[c]{@{}l@{}}CodeEditorBench\\ \textit{-switch-py}\end{tabular}~\cite{CodeEditorBench} &  \begin{tabular}[c]{@{}c@{}}Coding\\Challenge\end{tabular} & \begin{tabular}[c]{@{}l@{}}
\textbf{Description:} It is a curated set of coding challenges from various sources, including LeetCode. \\ 
\textbf{Reason:} This benchmark is leaked as the LLM pre-training data includes many repositories from\\ LeetCode. By examining the sources of pre-training data that duplicate this benchmark's data,\\ we identified 20 repositories whose names contain the string ``leetcode".\end{tabular} \\ 
\bottomrule
\end{tabular}
}
\vspace{-0.6cm}
\end{table*}

\vspace{0.2cm}
\noindent
\begin{tcolorbox} [boxrule=0.8pt,
                top=0.2pt,
                  bottom=0.2pt]
    \textbf{Answer to RQ2}: 
    \revised{
     We identified four potential causes of high data leakage: 1) the direct inclusion of benchmark data in the pre-training dataset, 2) overlap between the repositories used to create the benchmarks and the pre-training dataset, 3) reliance on coding platforms like LeetCode to construct the benchmarks, and 4) the shared use of data sources, such as GitHub issues, in both the benchmarks and the pre-training dataset. We also offer several actionable recommendations to reduce the risk of benchmark leakage in future studies.}
\end{tcolorbox}

\subsection{RQ3: Impact of Data Leakage for LLM Evaluation}

\noindent
\textbf{Motivation.}
This research question investigates whether data leakage leads to unfair evaluation of LLMs, specifically whether LLMs perform better on leaked data compared to non-leaked data. To address this, we conduct experiments using the APPS~\cite{hendrycks2021measuring} dataset, a widely used benchmark for code generation. We chose code generation as the focus of our analysis because it is a critical SE task where LLMs are frequently applied. We specifically selected the APPS dataset because it includes comprehensive test cases designed to evaluate the functional correctness of generated code. Moreover, the APPS dataset contains a sufficient number of leaked samples, as identified in our study, while many other code generation benchmarks lack enough leaked samples for a similar analysis.

\begin{table*}[b]
\centering
\vspace{-0.2cm}
\caption{Effectiveness of LLMs on Leaked and Non-Leaked Subsets of APPS}
\vspace{-0.2cm}
\label{table:rq3}
\scalebox{0.95}{
\rotatebox{0}{
\begin{tabular}{l|cll|cll}
\toprule
\multirow{2}{*}{\textbf{\begin{tabular}[c]{@{}l@{}}APPS\\ (zero-shot)\end{tabular}}} & \multicolumn{3}{c|}{\textbf{Leaked test samples}} & \multicolumn{3}{c}{\textbf{Non-leaked test samples}} \\ \cline{2-7} 
 & \multicolumn{1}{c|}{\textbf{pass@1}} & \multicolumn{1}{c|}{\textbf{pass@2}} & \multicolumn{1}{c|}{\textbf{pass@3}} & \multicolumn{1}{c|}{\textbf{pass@1}} & \multicolumn{1}{c|}{\textbf{pass@2}} & \multicolumn{1}{c}{\textbf{pass@3}} \\ 
 \toprule
\textbf{StarCoder-7b} & \multicolumn{1}{c|}{4.4\%} & \multicolumn{1}{c|}{8.9\%} & \multicolumn{1}{c|}{13.3\%} & \multicolumn{1}{c|}{0.9\%} & \multicolumn{1}{c|}{1.6\%} & \multicolumn{1}{c}{2.2\%} \\ \hline
\textbf{StarCoder-3b} & \multicolumn{1}{c|}{3.7\%} & \multicolumn{1}{c|}{7.4\%} &   \multicolumn{1}{c|}{11.1\%}  & \multicolumn{1}{c|}{0.4\%} & \multicolumn{1}{c|}{0.7\%} & \multicolumn{1}{c}{0.9\%}  \\ \hline
\textbf{StarCoder-1b} & \multicolumn{1}{c|}{3.7\%} & \multicolumn{1}{c|}{7.4\%} &   \multicolumn{1}{c|}{11.1\%} & \multicolumn{1}{c|}{0.3\%} & \multicolumn{1}{c|}{0.6\%} &  \multicolumn{1}{c}{0.9\%} \\ 
\bottomrule
\end{tabular}
}}
\end{table*}

\vspace{0.2cm}
\noindent
\textbf{Method and Setup.}
The APPS dataset consists of 10,000 code generation problems/data samples, each paired with Python solutions. These problems are categorized into three difficulty levels: introductory, interview, and competition, with solutions ranging from simple one-liners to complex algorithms. On average, each problem includes 21.2 test cases designed to assess the functional correctness of the generated code. The dataset is split into 5,000 samples for training and 5,000 samples for testing.

For our experiments, we first divide the APPS test set into two subsets: leaked and non-leaked. We then generate predictions for both subsets using LLMs of varying sizes: StarCoder-7b, StarCoder-3b, and StarCoder-1b. The models are prompted with a simple instruction format: \textit{``$\#\#\#$ Instruction: [instruction input from APPS data sample] $\#\#\#$ Response:''}, and the model generates a prediction accordingly. The models are prompted in a zero-shot manner, i.e., no labeled data is provided in the prompt. This approach allows us to capture the model's behavior without any influence from labeled data. In other words, the performance in a zero-shot setting more accurately reflects the model's direct outputs based on its pre-training.
For evaluation, we use the Pass@k metric, which is used in the original APPS paper. Pass@k measures the percentage of problems for which at least one of the top-k generated code samples can pass all test cases. This metric reflects the model's ability to produce fully correct solutions within k attempts.

\vspace{0.2cm}
\noindent
\textbf{Results.}
As shown in Table~\ref{table:rq3}, StarCoder-7b, StarCoder-3b, and StarCoder-1b consistently demonstrate superior performance on the leaked samples compared to non-leaked samples. Specifically, StarCoder-7b achieves Pass@1, Pass@2, and Pass@3 scores that are 4.9, 5.6, and 6.0 times higher, respectively, on the leaked samples compared to the corresponding metrics for the non-leaked samples. This indicates that data leakage results in an unfair evaluation of LLMs, with the models performing significantly better on leaked data than on non-leaked data.
In addition, the reason LLMs do not achieve perfect performance on the leaked samples could be attributed to the complexity of the benchmark data, which makes it difficult for the LLMs to fully memorize those data samples.

\vspace{0.2cm}
\noindent
\begin{tcolorbox} [boxrule=0.8pt,
                top=0.2pt,
                  bottom=0.2pt]
    \textbf{Answer to RQ3}: 
    \revised{
    The experimental results show that data leakage has a significant impact on the evaluation of LLMs, with models performing better on leaked data compared to non-leaked data. For example, StarCoder-7b achieves a Pass@1 score that is 4.9 times higher on the leaked samples than on the non-leaked samples. This underscores how the presence of leaked benchmark samples can lead to significantly inflated metrics, emphasizing the critical need to address the data leakage issue in SE benchmarks to ensure fair evaluation.}
\end{tcolorbox}

\subsection{RQ4: Automatic Data Leakage Detection Without Knowledge of Pre-training Data}

\noindent
\textbf{Motivation and Setup.}
Since pre-training data for many LLMs is inaccessible, another key question is whether SE benchmark leakage can be inferred solely from LLM behaviors, such as the LLM's confidence in its generated content, without knowledge of the LLM's pre-training data.

\begin{figure}[b] 
    \centering
    \vspace{-0.6cm}
    \includegraphics[width=10cm]{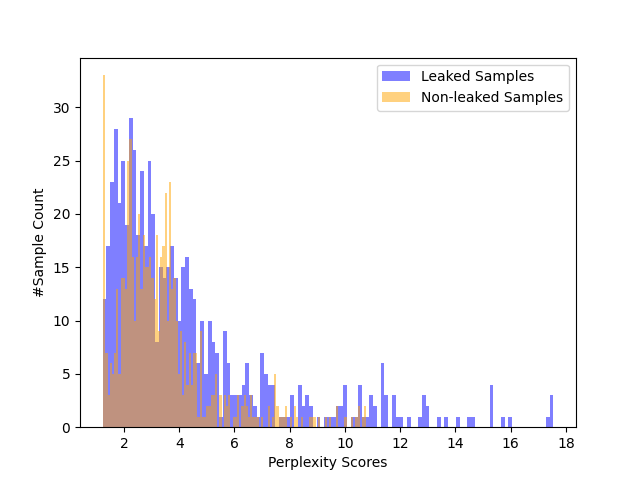} 
     \vspace{-0.3cm}
    \caption{Perplexity Distribution of Leaked and Non-leaked SE Benchmark Samples Using StarCoder-7B} 
    \label{fig:ppl}
\end{figure}

A benchmark dataset with reliable labels (i.e., leaked or non-leaked) is crucial for evaluating how effectively data leakage in LLMs can be detected when their pre-training data is unavailable. 
To address this, we utilize our manually labeled dataset created through our data leakage detection approach DetectLeak (as shown in Figure~\ref{fig:method_minhash}). In RQ1 and RQ2, we focus exclusively on leaked samples. In contrast, in RQ4, we include both leaked and non-leaked SE benchmark data samples to form a new dataset, named \textit{AutoDetectLeak-Bench}. Specifically, we extract the manually labeled leaked and non-leaked samples and perform a deduplication process. Since the dataset is imbalanced, we apply random under-sampling to equalize the number of leaked and non-leaked samples.
As a result, AutoDetectLeak-Bench consists of 1,300 samples and serves as a test set to evaluate the effectiveness of automated metrics in detecting data leakage.

\vspace{0.2cm}
\noindent
\textbf{Method.}
We investigate the automatic metric \textit{Perplexity}~\cite{Perplexity} to infer whether benchmark data has been leaked. Specifically, Perplexity measures the confidence of a language model when generating an output, with a lower Perplexity value indicating higher confidence in the model's predictions. The underlying assumption is that an LLM is more confident in examples it has encountered during pre-training. Therefore, intuitively, leaked data samples should exhibit lower Perplexity compared to non-leaked samples.

\vspace{0.2cm}
\noindent
\textbf{Results.}
Figure~\ref{fig:ppl} illustrates the Perplexity distribution of leaked and non-leaked SE benchmark samples, evaluated using the StarCoder-7b~\cite{starcoder_one} model. We removed the top 2\% of outliers (data samples with excessively high Perplexity) from both subsets to enhance the clarity of the visualization.
From Figure~\ref{fig:ppl}, we observe that the Perplexity of leaked samples is not necessarily lower than that of non-leaked samples. On the contrary, in the higher Perplexity range (e.g., Perplexity > 10), the number of leaked samples significantly exceeds that of non-leaked samples.

\begin{figure}[t] 
    \centering
    \includegraphics[width=10cm]{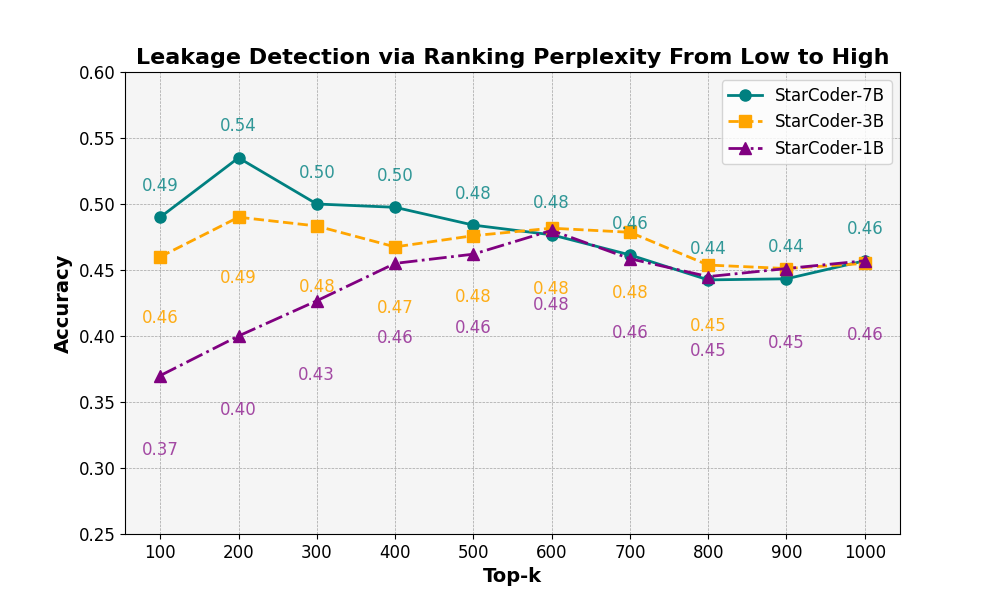} 
    \vspace{-0.3cm}
    \caption{Detection Accuracy (Proportion of Leaked Samples in Top-k) of Ranking Perplexity in Ascending Order} 
     \vspace{-0.7cm}
    \label{fig:ranking_ppl_accuracy}
\end{figure}

To quantitatively evaluate the effectiveness of using perplexity to infer benchmark data leakage, we rank all labeled samples in our dataset based on ascending Perplexity scores. The top-k samples, which have the lowest Perplexity, suggest higher LLM confidence and are hypothesized to be more likely leaked. We use a simple detection approach that classifies the top-k samples as leaked, while all remaining samples are classified as non-leaked. The value of k is a tunable hyperparameter.
Figure~\ref{fig:ranking_ppl_accuracy} presents the proportion of leaked samples (i.e., accuracy) within the top-k samples, ranked in ascending order, with results from LLMs of varying sizes: StarCoder-7b, StarCoder-3b, and StarCoder-1b. We evaluate multiple k-values, ranging from 100 to 1,000 in increments of 100.
As shown in Figure~\ref{fig:ranking_ppl_accuracy}, the accuracy of detecting leaked samples based on Perplexity ranking is relatively low across all three models.
Specifically, accuracy typically ranges between 40\% and 50\%, highlighting the challenges of using an automated metric like perplexity for reliable data leakage detection.

\vspace{0.2cm}
\noindent
\begin{tcolorbox} [boxrule=0.8pt,
                top=0.2pt,
                  bottom=0.2pt]
    \textbf{Answer to RQ4}: 
     Our experimental results indicate that detecting data leakage using an automated metric like Perplexity is challenging, with accuracy ranging from only 40\% to 50\% in most cases. We encourage the research community to explore more effective automated methods for identifying data leakage, especially in scenarios where the model's pre-training data is unknown.  
     Future research can use our manually labeled \textit{AutoDetectLeak-Bench} dataset to evaluate newly proposed methods for detecting data leakage when assuming the LLM pre-training data is not available.
\end{tcolorbox}

\section{Discussion}\label{section6}

\subsection{Implications}
Our study presents the first large-scale investigation into data leakage in software engineering benchmarks involving LLMs. Based on our findings, we highlight several implications for the SE research community to further explore and address the challenge of data leakage in future research.

\vspace{0.1cm}
\noindent
\textbf{Researchers need to be cautious about the potential data leakage in SE benchmarks.} Many LLMs are developed after the creation of widely used SE benchmarks, increasing the likelihood of data leakage. Such leakage can result in inflated performance metrics for LLMs, leading to unfair comparisons with non-LLM approaches. For example, our study shows that QuixBugs is entirely leaked to StarCoder, making it unsuitable for evaluating LLMs trained on StarCoder or its pre-training data. As a widely used open-source LLM family, StarCoder and its pre-training data serve as the foundation for many other LLMs, such as PanGu-Coder2~\cite{shen2023pangu}, WizardCoder~\cite{luo2023wizardcoder}, OctoPack~\cite{muennighoff2023octopack}, CodeShell~\cite{xie2024codeshell}, and DeepSeek-Coder~\cite{deepseekcoder}. Our findings on data leakage are applicable to these related LLMs as well.
While many researchers have been aware of this issue, the lack of dedicated studies and tools has often left them unable to verify whether their data was leaked. Our work introduces an effective framework, \textbf{DetectLeak}, to help mitigate the risk of data leakage in benchmark datasets. We believe that future researchers can leverage our framework to enhance the reliability of their LLM evaluations, particularly in addressing data leakage concerns. 
If the used LLMs do not disclose their pre-training data (like ChatGPT), researchers can instead perform leakage detection using the largest open-source LLM pre-training corpus available. While this approach cannot guarantee complete elimination of leakage, it can still significantly reduce the risk to the best extent possible.

\vspace{0.1cm}
\noindent
\textbf{Researchers are encouraged to use our cleaned version of SE benchmarks, LessLeak-Bench, in future studies.}
In this work, we identified many leaked benchmark samples that are no longer suitable for LLM evaluation. By removing these leaked samples, we created a cleaned version of the SE benchmarks, named \textbf{LessLeak-Bench}. For researchers currently using or planning to use the benchmarks analyzed in our study, we recommend adopting LessLeak-Bench to ensure more reliable and trustworthy evaluation results.

\vspace{0.1cm}
\noindent
\textbf{Developing more advanced methods to effectively detect leaked benchmark samples without access to LLM pre-training data remains an essential need.} In our study, we explored the use of Perplexity scores to detect leaked samples (RQ4). While this approach showed potential, there remains significant room for improvement. To support progress in this area, we introduce the \textbf{AutoDetectLeak-Bench} dataset, which contains manually verified labels for both leaked and non-leaked samples. This benchmark serves as a foundation for future research aimed at developing more effective leakage detection methods, particularly in scenarios where LLM pre-training data is unavailable.
As the first large-scale study on data leakage in SE benchmarks, our focus was on LLMs with publicly available pre-training data. This choice was necessary due to the current lack of highly effective automated tools capable of accurately identifying data leakage without knowledge of pre-training data in the SE community. To ensure the reliability of our findings, we selected LLMs that openly shared their pre-training data. Using this information, we constructed AutoDetectLeak-Bench.
Before this work, the absence of a reliable, manually verified dataset made it challenging to assess the effectiveness of automated leakage detection tools. By providing AutoDetectLeak-Bench, we address this gap and offer a valuable resource for advancing research in this area. This dataset can support the development of accurate leakage detection methods, facilitating the identification and removal of leaked data even for models like Llama3 that do not disclose their pretraining data. Such advancements will help ensure more robust and fair evaluations in the future.

\subsection{Actionable Suggestions}
\subsubsection{\textbf{Suggestions for LLM Providers}}
LLM providers like OpenAI and Meta have developed exceptional and powerful models. To address potential data leakage issues, a key measure is for providers to carefully curate their pre-training data, ensuring it does not contain duplicates or overlaps with SE benchmark samples. 

While some LLM providers such as Huggingface~\cite{starcoder_one} have performed de-duplication for well-known SE benchmarks such as HumanEval and MBPP, many other SE benchmarks have been overlooked in this process.
We recommend that LLM providers adopt the following practices:
\begin{itemize}[leftmargin=*]
    \item Remove pre-training data samples that overlap with a wider range of SE benchmark datasets, rather than limiting removal to popular code generation benchmarks like HumanEval and MBPP. For example, LLM providers could consider the studied benchmarks in this work. 
    \item Carefully inspect GitHub repositories used as part of the pre-training data. If a repository's name or README file explicitly indicates that it was cloned from a benchmark or matches the name of a known benchmark, providers should avoid including such repositories in their pre-training data. This precaution helps prevent severe data leakage, especially when entire cloned versions of benchmark data are used in pre-training.
    \item Be cautious about repositories related to coding platforms like LeetCode. Additionally, repositories containing common coding website names, such as ``leetcode,'' or algorithm-related keywords like ``data-structures,'' should be excluded. Many benchmarks are sourced from these platforms or algorithm-related repositories, which increases the likelihood of data leakage.
    \item Consider leveraging our approach to identify and remove duplicates. Utilize our methodology to detect and exclude pre-training data samples that overlap with the expanded set of SE benchmarks, ensuring cleaner and more reliable pre-training datasets.
    \item Apply leakage detection to instruction-tuning datasets as well. Instruction-tuning datasets, often generated by stronger LLMs to enhance other models, may also introduce data leakage. These datasets should undergo thorough checks to identify and remove overlapping samples with SE benchmarks to prevent inadvertent leakage.
\end{itemize}

\subsubsection{\textbf{Suggestions for LLM Users}}
Many LLMs have already been developed and may not have addressed data leakage from various SE benchmarks. Re-training these models to fix data leakage issues can be both costly and impractical. A more cost-effective approach is for LLM users to remove the leaked benchmark samples, thus reducing the risk of data leakage without the need for re-training the models.
Based on our findings in this study, we offer the following suggestions for LLM users:

\begin{itemize}[leftmargin=*]
    \item \textbf{Existing Benchmark Usage:} 
    For benchmarks that are included in LessLeak-Bench, it is recommended to prioritize using the cleaned versions to reduce the risk of data leakage. For benchmarks that are not included, researchers can follow our DetectLeak approach to identify and remove any potential leaked samples.

    \item \textbf{New Benchmark Construction:} 
    We recommend being cautious about using coding websites like LeetCode or algorithm-focused tutorial repositories to construct new benchmarks. These sources are commonly included in LLM pre-training datasets, which increases the risk of data leakage.   
    One potential solution is to incorporate new coding tasks introduced after the release of the LLMs on coding websites. For example, Liu et al.~\cite{DBLP:journals/tosem/LiuLWTLLL24} found that ChatGPT’s performance drops by up to five times on LeetCode coding tasks introduced after January 2022, likely because these new tasks are less susceptible to leakage in LLM training data.

    \item \textbf{Community Effort:}
    With the vast number of SE benchmarks available for studying various SE tasks, it is impractical for a single research group to examine data leakage across all of them. This paper provides an analysis of the data leakage status for a relatively large set of benchmarks. We encourage the SE community to build upon our work, such as by using our framework to identify and eliminate leaked samples in more SE benchmarks. Furthermore, we hope this initiative will raise greater awareness of data leakage issues and motivate the development of more sophisticated detection tools, regardless of whether access to LLMs’ pre-training data is available.
\end{itemize}

\subsection{Threats to Validity}
Our findings are limited to the specific SE benchmarks and LLMs studied, and therefore, may not generalize to all SE benchmarks and LLMs. To mitigate this limitation, we selected 83 diverse SE benchmarks to cover a wide range of popular software engineering tasks, including code generation, program repair, vulnerability detection, and more. For the LLMs, we chose StarCoder~\cite{starcoder_one}, a representative model that is fully open-sourced, including its model parameters and pre-training data.
We consider StarCoder a representative choice for this study because it demonstrates competitive performance among fully open-source models and serves as the foundation for derivative LLMs such as WizardCoder~\cite{luo2023wizardcoder}, OctoPack~\cite{muennighoff2023octopack}, CodeShell~\cite{xie2024codeshell}, and DeepSeek-Coder~\cite{deepseekcoder}.
Due to the immense size of LLM pre-training data, manual verification of data leakage is impractical. Instead, we adopted an efficient automated technique to identify potentially leaked data, followed by manual labeling of the flagged samples to confirm leaked SE benchmarks. However, we acknowledge that this approach may not detect all leaked SE benchmark samples, as the automated technique could miss some cases. Despite this limitation, our study identified a substantial number of leaked samples, which can help researchers avoid using these confirmed leaked samples for evaluation. Additionally, our findings provide an overview of the current state of benchmark leakage within the SE community.
To further support the community, we have offered actionable recommendations for both LLM providers and users and shared the implications of our work.

\section{Related Work}\label{section7}

Yang et al.~\cite{DBLP:conf/icse/0003ZWS0H024} investigates memorization in Code LLMs, focusing on CodeParrot and CodeParrot-small. The authors use code clone detection tools to identify duplicates in model-generated outputs and analyze patterns of memorization. Our work differs in three key ways. First, we focus on the StarCoder series, a larger and more advanced family of Code LLMs. Second, while Yang et al.~\cite{DBLP:conf/icse/0003ZWS0H024} examines general memorization capabilities using pre-training data not specifically tied to SE benchmarks, our study directly addresses data leakage in SE benchmarks, offering insights into the risks and biases such leakage introduces. Third, unlike their reliance on clone detection tools, which can yield false positives, our approach incorporates large-scale manual labeling to ensure the accuracy and reliability of identified duplicates and leaked samples.

Lopez et al.~\cite{lópez2024interdatasetcodeduplicationdata} investigated dataset leakage in lightweight LLMs, such as CodeBERT, by leveraging an automatic clone detection tool to identify duplicates between pre-training data and three SE benchmarks.
Compared to Lopez et al., our study investigates larger and more advanced LLMs, such as StarCoder, which are approximately 125 times larger than the models analyzed by Lopez et al. and significantly more capable in generation tasks like code generation and program repair. Additionally, rather than focusing on only three SE benchmarks, our work evaluates a much broader set of 83 SE benchmarks across three popular programming languages, providing a more comprehensive analysis of benchmark leakage. To ensure the accuracy of identified leakage, our study also integrates large-scale manual labeling.

In addition to academic literature, several studies in the gray (non-peer-reviewed) literature have explored related topics.
Matton et al.~\cite{matton2024leakagecodegenerationevaluation} analyzed potential data leakage in the HumanEval and MBPP benchmarks by checking for occurrences of their prompts in public GitHub repositories. Similarly, Riddell et al.~\cite{riddell2024quantifyingcontaminationevaluatingcode} studied leakage in these benchmarks using a multi-step approach. They calculated Levenshtein similarity scores to identify overlaps between the benchmarks and LLM pre-training data, selected the top 500 pre-training samples with the highest similarity for each test case, and used an automatic code plagiarism detection tool to identify duplicates. 
In contrast to these gray literature studies, which focus on only two or three SE benchmarks, our work evaluates a significantly broader set of 83 SE benchmarks across three popular programming languages, enabling a more comprehensive analysis of benchmark leakage. Furthermore, our approach incorporates large-scale manual labeling to enhance the accuracy and reliability of leakage detection. We also provide actionable recommendations for LLM developers and users to effectively mitigate data leakage.

\section{Conclusion and Future Work}\label{section8}

In this study, we performed the first large-scale analysis of data leakage across 83 software engineering (SE) benchmarks, covering three popular programming languages—Python, Java, and C/C++. By combining an efficient near-duplicate detection algorithm with extensive manual labeling, we ensured the accurate identification of leaked data.

Our findings show that while data leakage is generally low, with average leakage ratios of 4.8\%, 2.8\%, and 0.7\% for Python, Java, and C/C++ benchmarks respectively, some benchmarks exhibit higher leakage that requires attention. We identified four main causes of leakage: direct inclusion of benchmark data in pre-training datasets, overlap between source repositories, reliance on platforms like LeetCode, and shared data sources such as GitHub issues.
We also found that automatic detection methods, like Perplexity-based metrics, struggle to distinguish between leaked and non-leaked samples. Additionally, our experiments reveal that data leakage inflates evaluation metrics, with models performing significantly better on leaked samples. For instance, StarCoder-7b achieved a Pass@1 score 4.9 times higher on leaked samples, underlining the need to address leakage to ensure fair evaluations.
This study offers insights into data leakage status in SE benchmarks and its impact on LLM evaluation.

In the future, we aim to expand the analysis to additional benchmarks and explore new methods to prevent or further reduce data leakage.

\vspace{0.2cm}
\noindent \textbf{Acknowledgement.}  This research / project is supported by the National Research Foundation, under its Investigatorship Grant (NRF-NRFI08-2022-0002). Any opinions, findings and conclusions or recommendations expressed in this material are those of the author(s) and do not reflect the views of National Research Foundation, Singapore.

\balance
\bibliographystyle{acm}
\bibliography{main}

\end{document}